\documentclass[useAMS,usenatbib,usegraphicx]{mn2e}
\usepackage{amssymb}
\usepackage[british]{babel}
\title[Radio spectra and polarisation of BAL QSOs]{Radio spectra and polarisation properties of radio-loud Broad Absorption Line Quasars}

\author[F. M. Montenegro-Montes et al.]
          {F. M. Montenegro-Montes,$^{1,2}$\thanks{E-mail:fmm@ira.inaf.it} 
           K.-H. Mack,$^1$ M. Vigotti,$^1$ 
	   C. R. Benn,$^3$ R. Carballo,$^4$ 
\newauthor J. I. Gonz\'alez-Serrano,$^5$ 
           J. Holt,$^6$ F. Jim\'enez-Luj\'an$^{5,7}$\\
$^{1}$Istituto di Radioastronomia, INAF, Via Gobetti 101, I-40129 Bologna, Italy\\
$^{2}$Departamento de Astrof\'isica. Universidad de La Laguna, Avda. Astrof\'isico Fco. S\'anchez s/n, E-38200 La Laguna, Spain.\\
$^{3}$Isaac Newton Group, Apartado 321, E-38700 Santa Cruz de La Palma, Spain\\
$^{4}$Dpto. de Matem\'atica Aplicada y Ciencias de la Computaci\'on. Univ. de Cantabria, ETS Ingenieros de Caminos
Canales y Puertos.\\ Avda. de los Castros s/n. E-39005 Santander, Spain\\
$^{5}$Instituto de F\'isica de Cantabria (CSIC-Universidad de Cantabria), Avda. de los Castros s/n, E-39005 Santander, Spain\\
$^{6}$Leiden Observatory. Leiden University, P O Box 9513, NL-2300 RA Leiden, The Netherlands\\
$^{7}$Dpto. de F\'isica Moderna,  Univ. de  Cantabria, Avda de los Castros s/n,  E-39005 Santander, Spain\\}
\begin{document}

\date{June, 2008, Draft 7}

\pagerange{\pageref{firstpage}--\pageref{lastpage}} \pubyear{2008}

\maketitle

\label{firstpage}

\begin{abstract}
We present multi-frequency observations of a sample of 15 radio-emitting Broad Absorption Line Quasars 
(BAL QSOs), covering a spectral range between 74 MHz and 43 GHz. They display mostly convex radio spectra 
which typically peak at about 1-5 GHz (in the observer's rest-frame), flatten at MHz frequencies, probably 
due to synchrotron self-absorption, and become steeper at high frequencies, i.e., $\nu \gtrsim$20 GHz.
VLA 22-GHz maps (HPBW $\sim$ 80 mas) show unresolved or very compact sources, with linear projected sizes of 
$\leq$1 kpc. About 2/3 of the sample look unpolarised or weakly polarised at 8.4 GHz, frequency in which 
reasonable upper limits could be obtained for polarised intensity. Statistical comparisons have been made 
between the spectral index distributions of samples of BAL and non-BAL QSOs, both in the observed and the 
rest-frame, finding steeper spectra among non-BAL QSOs. However constraining this comparison to compact 
sources results in no significant differences between both distributions. This comparison is consistent
with BAL QSOs not being oriented along a particular line of sight. In addition, our analysis of the
spectral shape, variability and polarisation properties shows that radio BAL QSOs share several properties 
common to young radio sources like Compact Steep Spectrum (CSS) or Gigahertz-Peaked Spectrum (GPS) sources.
\end{abstract}


\section{Introduction}

About 15 per cent of optically-selected quasars exhibit Broad Absorption Lines (BALs) in the blue wings of the 
UV resonance lines, 
due to gas with outflow velocities up to 0.2 c \citep{Hewett}. A classification of BALs has been done 
according to their association with highly ionised species like C\,{\sc iv}, N\,{\sc v}, Si\,{\sc iv} (HiBALs) or, 
in addition, with
low-ionisation species like Mg\,{\sc ii} or Al\,{\sc iii} (LoBALs). These broad troughs are the most evident
manifestations of quasar outflows but many other intrinsic absorbers such as Associated Absorption Lines (AALs), 
mini-BALs or high velocity Narrow Absorption Lines (NALs) are also related to them (\citealt{Hamann};
\citealt{Ganguly2}).

The study of BAL quasars (BAL QSOs) has become increasingly important in the last years. Outflows are not only 
useful as probes of the physics in the AGN environment, but also because they might be an important piece
in many other puzzles like enrichment of the intergalactic medium, galaxy formation, evolution of the AGN and
the host galaxy, cluster cooling flows, magnetisation of cluster and galactic gas and the luminosity function 
of quasars. 

At the moment it is not clear how many quasars host outflows and why these can only be seen in a fraction of 
the quasar population. The most popular hypotheses proposed to explain this mainly differ in the role given 
to orientation. The main premise in the `orientation scenario' is that the BAL phenomenon might 
be present in all quasars but intercepted by only $\sim$15 per cent of the lines of sight to the quasar \citep{Weymann}, 
e.g. within the walls of a bi-funnel centred on the nucleus \citep{Elvis}. This scenario is supported because
in most respects, apart from the BALs themselves, BAL QSOs look similar to normal quasars. The small differences 
from non-BAL QSOs, e.g. slightly redder continua \citep{Reichard2} and higher optical polarisation, could be a 
consequence of a preferred orientation of viewing angle. 

The alternative hypothesis, the so-called `unification by time', assumes that BALs appear during one or maybe 
more short periods in the lifetime of quasars. The observed rate of BAL QSOs would then be a function of the 
percentage of the quasar lifetime in which these outflows show up. This evolutionary 
scenario has been discussed by \cite{Becker1} on the basis of the radio properties of their Faint Images Radio Sky 
at Twenty-centimeters (FIRST, \citealt{Becker95}) BAL QSO sample. They note that 80 per cent of their BAL QSOs are 
unresolved at a 0.2-arcsec scale, with both flat and steep spectra. This latter group resembles the Compact 
Steep-Spectrum Sources (CSS) and their Gigahertz Peaked-Spectrum (GPS) sub-class. As a consequence, they suggest 
the picture in which BAL QSOs represent an early stage in the development of quasars. 

This last scenario finds motivation on the collection of theoretical and phenomenological works that in recent years 
have contributed to build a first consistent picture of early AGN evolution. As an example, recent simulations 
\citep{Hopkins} describe how quasar evolution can be understood as the result of merger events, where the super-massive 
black hole grows via accretion and, when it finally becomes an active `protoquasar', expels out the surrounding material 
deposited by the merger through powerful winds. The fingerprints of these winds have been manifested not only as BAL
troughs in UV spectra, but also as emission line outflows and sometimes associated to H\,{\sc I} absorptions (see e.g., 
\citealt{Holt2}). To that respect, even if the fraction of optically-selected quasars hosting BAL systems is estimated to 
be about 10-20 per cent of the quasar population, it has been argued that these samples might be substantially biased against 
BAL QSOs, as suggested by larger fractions of BAL QSOs found among infrared-selected quasars \citep{Dai}. In addition, 
\cite{Ganguly2} have recently estimated that $\sim$60 per cent of AGNs could host outflows when the different kinds of intrinsic 
absorptions are considered. 

Radio emission has become, in fact, an important additional diagnostic tool when studying the orientation and 
evolutionary status of BAL QSOs. For a long time radio-loud BAL QSOs were believed to be extremely rare among the 
population of luminous quasars \citep{Stocke}. With the advent of large comprehensive radio surveys it has however 
become clear that radio-loud BAL QSOs are common. However the fraction of BAL QSOs seem to vary inversely with the 
radio-loudness parameter, with the radio-brightest quasars being 4 times less likely to exhibit BALs (\citealt{Becker2};
\citealt{Gregg2}). The evolutionary picture is, in fact, supported by some recent works coming from radio observations. 
For instance, EVN observations at 1.6 GHz of a few BAL QSOs from Becker's list \citep{Jiang} show that they present a 
still compact structure at these resolutions, with sub-kpc projected sizes, and a variety of orientations according to 
their radio geometry. In addition, the radio powerful CSS BAL QSO 1045+352 has also been observed with MERLIN and VLBA, 
and its complex radio structure has been interpreted as evidence of radio-intermittent activity \citep{Magda}.

Some other works address the question of the orientation of BAL QSOs through radio variability (\citealt{Zhou}; 
\citealt{Ghosh}). They compare the 1.4-GHz flux densities of all quasars in the Sloan Digital Sky Survey (SDSS) quasar 
catalogues (\citealt{Schneider05}; \citealt{Schneider07}) with detections in both the FIRST and NRAO VLA Sky Survey 
(NVSS, \citealt{Condon}) surveys. The most strongly variable quasars have probably their jets closely aligned with the 
line of sight, being the relativistic beaming responsible for such high flux density variations. Some of those  
variable quasars present in addition the BAL phenomenon which is considered as a proof for the existence of at least 
a sub-population  of polar BAL QSOs. At the same time, a few examples have been found of BAL QSOs associated to radio 
sources with an extended FR II morphology \citep{Gregg2} which suggest for those the opposite edge-on orientation. 

This paper reports the first results of a study in which a systematic approach is taken to characterise the radio-loud 
BAL QSO population. The final goal is to locate these objects in the framework of an adequate orientation and/or evolutionary 
status scenario. In order to do this, the radio spectra and polarisation properties of a small sample of radio-loud BAL QSOs 
will be presented.

In Section \ref{sec2} some samples of radio emitting QSOs existing in the literature are summarised, and among them a small
sample of BAL QSOs is selected for multi-frequency continuum and polarisation observations. In Section \ref{sec3} the
observations, data reduction and the treatment of the errors are described. In Section \ref{sec4} the results of the 
observations are presented. The radio morphologies of BAL QSOs, their polarisation properties and the variability of
some of these objects with 2-epoch observations are analysed. In addition, the shape of the radio spectra is described
and the radio spectral index distributions of radio quasars with and without associated BALs are compared. All these 
results are reviewed together in Section \ref{sec6} and compared with other findings trying to put emphasis on those 
aspects related to the orientation vs. evolutionary dilemma. Some of the limitations of the analysis are noted and 
solutions to some of these caveats are proposed. Finally Section \ref{sec7} summarises the main conclusions. 

The cosmology adopted within the paper assumes a flat universe and the following parameters: $H_{0}$=70 km s$^{-1}$ 
Mpc$^{-1}$, $\Omega_{\Lambda}$=0.7, $\Omega_{M}$=0.3. The adopted convention for the spectral index definition is 
$S_{\nu} \propto \nu^{\alpha}$, except where otherwise specified.

\begin{table*}
 \centering
 \begin{minipage}{152mm}
  \caption{Sample of 15 radio-loud BAL QSOs studied in this paper. 
  }\label{listsample}
  \begin{tabular}{@{}cccccrccccc@{}}
  \hline
   \multicolumn{1}{c}{ID}           &
   \multicolumn{1}{c}{RA}           & 
   \multicolumn{1}{c}{DEC}          & 
   \multicolumn{1}{c}{$r_{ro}$}     &
   \multicolumn{1}{c}{z}            &  
   \multicolumn{2}{c}{$S_{\rm 1.4 ~GHz}^{peak}$} & 
   \multicolumn{1}{c}{E}            & 
   \multicolumn{1}{c}{log($L_{\rm 5 ~GHz}$)} & Ref & Type\\
           & (J2000)  & (J2000) &  (arcsec) &             & \multicolumn{2}{c}{(mJy/beam)}  &       & (W Hz$^{-1}$) &     &       \\
      (1)  & (2)      & (3)     &  (4)      &      (5)    & \multicolumn{2}{c}{(6)}  &  (7)  &  (8)   & (9) &  (10) \\
\hline
0039$-$00  & 00:39:23.18 & $-$00:14:52.6 & 0.22 & 2.233 & 21.2 & & 19.50 & 26.33 & 1 & HiBAL \\
0135$-$02  & 01:35:15.22 & $-$02:13:49.0 & 0.31 & 1.820 & 22.4 & & 16.79 & 26.22 & 3 & LoBAL \\
0256$-$01  & 02:56:25.56 & $-$01:19:12.1 & 0.66 & 2.491 & 27.6 & & 18.57 & 26.54 & 3 & HiBAL \\
0728+40    & 07:28:31.64 &   +40:26:16.0 & 0.33 & 0.656 & 17.0 & & 15.27 & 24.97 & 2 & LoBAL \\
0837+36    & 08:37:49.59 &   +36:41:45.4 & 0.20 & 3.416 & 25.5 & & 19.16 & 26.66 & 6 & LoBAL \\
0957+23    & 09:57:07.37 &   +23:56:25.2 & 0.12 & 1.995 &136.1 & & 17.64 & 27.05 & 2 & HiBAL \\
1053$-$00  & 10:53:52.86 & $-$00:58:52.7 & 0.14 & 1.550 & 24.7 & & 18.03 & 26.02 & 4 & LoBAL \\
1159+01    & 11:59:44.82 &   +01:12:06.9 & 0.07 & 1.989 &268.4 & & 17.30 & 27.31 & 1 & HiBAL \\
1213+01    & 12:13:23.94 &   +01:04:14.7 & 0.13 & 2.836 & 21.5 & & 19.69 & 26.65 & 4 & HiBAL \\
1228$-$01  & 12:28:48.21 & $-$01:04:14.5 & 0.27 & 2.653 & 29.4 & & 17.74 & 26.62 & 4 & HiBAL \\
1312+23    & 13:12:13.57 &   +23:19:58.6 & 0.10 & 1.508 & 43.3 & & 17.13 & 26.32 & 2 & HiBAL \\
1413+42    & 14:13:34.38 &   +42:12:01.7 & 0.10 & 2.810 & 17.8 & & 17.63 & 26.42 & 2 & HiBAL \\
1603+30    & 16:03:54.15 &   +30:02:08.6 & 0.28 & 2.028 & 53.7 & & 17.61 & 26.61 & 2 & HiBAL \\
1624+37    & 16:24:53.47 &   +37:58:06.6 & 0.05 & 3.377 & 56.1 & & 17.62 & 27.06 & 5 & HiBAL \\
1625+48    & 16:25:59.90 &   +48:58:17.5 & 0.02 & 2.724 & 25.3 & & 17.41 & 26.59 & 6 & HiBAL \\
\hline
\end{tabular}
Columns are: (1) source ID; (2,3) Optical 
coordinates in J2000.0; (4) angular distance between the optical and FIRST radio positions; (5) optical spectroscopic 
redshift; (6) FIRST 1.4-GHz flux density; (7) APS E magnitude corrected from galactic extinction; (8) 5-GHz rest-frame 
luminosity, computed using the radio spectra presented in this paper; (9) reference where the BAL QSO appears in the 
literature and (10) BAL classification as given in that reference.\\
References key: 1- \cite{Menou}; 2- \cite{Becker1}; 3- \cite{Becker2};  4- \cite{Reichard}; 5- \cite{Holt}; 
6- Identified directly from SDSS-DR3 database. 
\end{minipage}
\end{table*}

\section[]{Samples of BAL and non-BAL QSOs used}\label{sec2}

The first significant sample of radio BAL QSOs was presented by \cite{Becker1}, which included 29 BAL QSOs. 
These were optical identifications of FIRST radio sources in the FIRST Bright Quasar Survey (FBQS) showing BAL features in 
their optical spectra. \cite{Becker1} performed radio observations of this list at two frequencies, 3.6 and 21 cm, in order to 
determine the radio spectral index of these sources. 

One of the aims of our work is to present a representative sample of radio-loud BAL QSOs with radio spectra covering a 
wide range in frequency (i.e. between 74 MHz and 43 GHz). The selection was based on the available samples of radio-loud 
BAL QSOs in the literature when the project was started. The main selection criterion was a cut in flux density of 
$S_{\rm{1.4~GHz}} >$ 15 mJy in order to facilitate total power detections at high frequencies and also the detection of polarised 
emission from those sources with a higher degree of linear polarisation.

\begin{figure}
  \includegraphics[width=92mm]{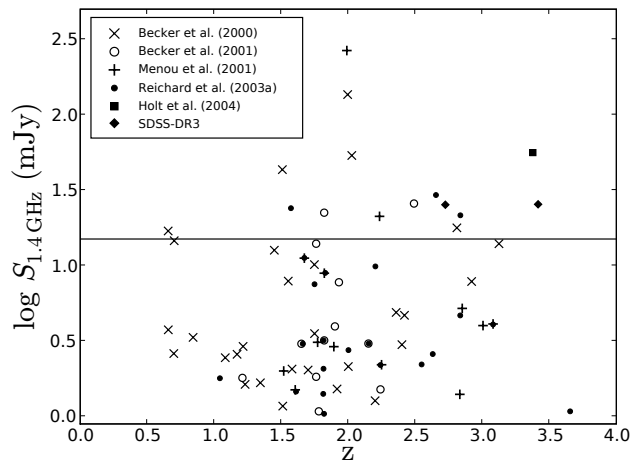}
  \caption{Distribution of known radio-loud BAL QSOs in log S and z before going on-sky. The sample to be studied in
  this paper was selected by taking all those with $S_{1.4 \rm{GHz}} >$ 15 mJy. 
  }\label{flux_vs_z_v2}
\end{figure}

We have defined a sample of radio-loud BAL QSOs [RBQ sample] composed of 15 sources. Seven of them were present 
in the list of \cite{Becker1} or its extension in the southern galactic cap \citep{Becker2}. The remaining eight 
objects are SDSS quasars classified as BAL QSOs by visual inspection and associated to FIRST sources within a 
2$\arcsec$ radius. Two of these were presented by \cite{Menou}, three more by \cite{Reichard} and the unusual 
BAL QSO 1624+37 in the sample was discovered in a survey for z$\sim$4 quasars \citep{Holt}. These 13 objects 
are \emph{the only} BAL QSOs present in the mentioned samples with $S_{\rm{1.4~GHz}} >$ 15 mJy. Finally, the other 
two BAL QSOs satisfying this flux density criterion were identified by one of us directly from the third data 
release of the SDSS (DR3). Table \ref{listsample} shows the main properties of the RBQ sample. Figure 
\ref{flux_vs_z_v2} shows flux density at 1.4 GHz versus redshift for all BAL QSOs belonging to the mentioned 
samples. Sources above the horizontal line of 15 mJy define the RBQ sample.

At a later stage, \cite{Trump} presented a catalogue with 4784 SDSS BAL QSOs selected using a more homogeneous 
selection criteria. They used an automated algorithm to fit the continuum and measure the AI (Absorption Index, 
\citealt{Hall}) in all quasars from the SDSS-DR3 quasar catalogue \citep{Schneider05}. In fact, all SDSS quasars in the 
RBQ sample belong to the list of \cite{Trump}, except 0039-00 which is not included in the SDSS-DR3 quasar catalogue, but appears 
in the more recent SDSS-DR5 quasar catalogue \citep{Schneider07}. However, some care has to be taken when using the list
of \cite{Trump}. According to the more subjective classification by \cite{Ganguly}, the absorption present in about 15 per cent
of the quasars classified as BAL QSOs in Trump's list might actually be due to intervening systems instead of having an intrinsic
origin. Some examples of these false positives are shown in Figure 2 of \cite{Ganguly}.

In this work the radio spectra of BAL QSOs will be described, and in particular, the spectral index distribution will be 
compared with that of a non-BAL QSO sample. This is an interesting exercise because the spectral index is known to be
an statistital indicator of the orientation of quasars, and this comparison can give some information about the orientation 
of BAL QSOs. This comparison sample of non-BAL QSOs has been extracted from the complete B3-VLA quasar catalogue \citep{Vigotti97}. 
The B3-VLA survey \citep{Vigotti89} consists of 1049 sources selected from the B3 survey \citep{Ficarra} in five flux density 
bins between declination 37 and 47 deg. It was designed to be complete down to flux densities of 100 mJy at 408 
MHz. In addition to the original flux densities at 408 MHz the entire classical radio frequency range has been covered for 
most of the sources in the sample, both from already existent surveys (6C at 151 MHz, WENSS at 325 MHz, NVSS at 1400 MHz, 
GB6 at 4850 MHz) and from dedicated observations, mainly at higher frequencies (74 MHz, \citealt{Mack}; 2.7 GHz, \citealt{Klein}; 
4.85 GHz, \citealt{Vigotti99} and 10.5 GHz, \citealt{Gregorini}). Based on this radio survey, \cite{Vigotti97} presented a 
complete sample of 125\footnote{In practise 124 because the redshift of quasar B3 2329+398 is not known.} quasars and 
this will be used as a comparison sample of radio-loud non-BAL quasars.

\begin{figure}
  \includegraphics[width=92mm]{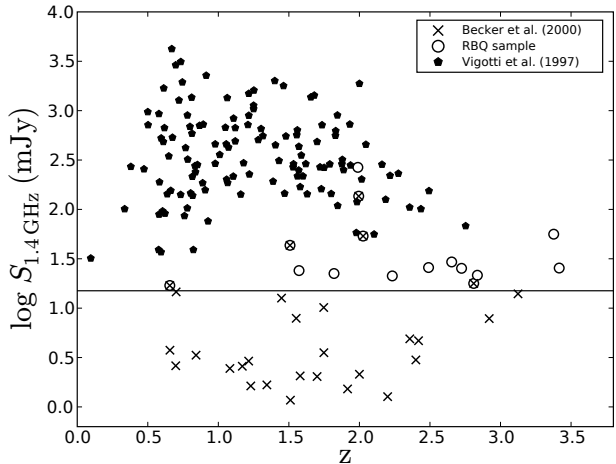}
  \caption{Same plot as Figure \ref{flux_vs_z_v2} in which the sample of normal radio-loud quasars from Vigotti et 
  al. (1997) is shown, together with the samples of radio-fainter BAL QSOs.}\label{flux_vs_z_v3}
\end{figure}

In Figure \ref{flux_vs_z_v3} the flux density is plotted against the redshift but now the sample of \cite{Vigotti97} 
is presented together with that of \cite{Becker1} and the one studied in this paper. The redshift range covered by the BAL QSOs 
goes from 0.656 to 3.124 (Becker's list) and to 3.416 (RBQ sample). The sources in \cite{Vigotti97} list are, on average at 
smaller redshift ($z_{mean}$=1.25), but they cover the redshift range between 0.096 and 2.757. Figure \ref{flux_vs_z_v3} also 
illustrates how the RBQ sample represent intermediate flux densities, between the flux density coverage spanned by the relatively 
faint sample of \citep{Becker1} and that covered by the list of brighter quasars from \cite{Vigotti97}.

\section{Observations and data reduction}\label{sec3}

Radio-continuum flux densities with full polarisation information were collected for the RBQ sample at different frequencies 
in several runs with the Effelsberg 100-m radiotelescope and the Very Large Array (VLA). A summary of the different 
observing runs can be found in Table \ref{table_runs}, and the list of frequencies used with their characteristics (e.g., 
angular resolutions) are displayed in Table \ref{observations}.

\begin{table}
 \centering
 \begin{minipage}{140mm}
  \caption{Summary of the observations.}\label{table_runs}
  \begin{tabular}{clcc}
  \hline
     Run      &   Date  &  Telescope  &  Frequencies (GHz)\\
  \hline
1& 26$-$30 Jan 05  & Effelsberg &4.85, 8.35, 10.5\\
2& 15$-$16 Jul 05  & Effelsberg &4.85, 8.35, 10.5\\
3& 27 Oct 05       & Effelsberg &4.85, 8.35, 10.5\\
4& 23$-$26 Jun 06  & Effelsberg &2.65\\
5& 13 Feb 06       & VLA(A)     &8.45, 14.5, 22.5, 43.5\\
6&  7 Mar 06       & VLA(A)     &8.45, 14.5, 22.5, 43.5\\
\hline
\end{tabular}
\end{minipage}
\end{table}

\subsection{Effelsberg 100-m telescope}

The single-dish observations at 4.8, 8.4 and 10.5 GHz took place in January 2005. Some sources were re-observed in July 
and October 2005 in order to confirm doubtful values or improve the signal-to-noise ratio. Observations at 2.65 GHz were
carried out in June 2006. All receivers used in these observations are mounted on the secondary focus of the 100-m antenna.
The recently installed 2.7-GHz hybrid feed can observe simultaneously in 9 channels, 8 of them 10-MHz wide and the 
ninth covering the full 80-MHz range. Since the sources in our sample are too faint to have a high signal-to-noise 
ratio per channel, only the wide-band channel centred at 2.65 GHz has been used. The receivers at 4.8 and 
10.5 GHz have multi-feed capabilities with 2 and 4 horns, respectively, allowing real-time sky subtraction in 
every subscan measurement. 

The observational strategy consisted of cross-scanning the source position in azimuth and elevation. These perpendicular 
scans were added to gain a factor $\sqrt{2}$ in sensitivity since the higher resolution FIRST maps ($\sim$ 5$\arcsec$) 
showed that none of these sources has structure more extended than 1 arcmin, the higher resolution in our observations 
(see Table \ref{observations}). The cross-scan length was chosen to be 
about four times the beam size at each frequency for a correct subtraction of linear baselines. The scanning speed 
ranged from 8$\arcsec$ to 32$\arcsec$/s depending on the frequency, stacking typically 8 subscans at 2.65 GHz, and 
some 16 subscans in the rest of the frequencies (32 for the faintest sources at 10.5 GHz). This translates into on-source 
integration times between 15-75 sec per source and per frequency. Before combining, individual subscans were checked 
to remove those affected by radio frequency interference, bad weather, or detector instabilities.

The calibration sources 3C\,286 and 3C\,295 were regularly observed to correct for both time-dependent gain 
instabilities and elevation-dependent sensitivity of the antenna and to bring our measurements to an absolute 
flux density scale \citep{Baars}. The quasar 3C\,286 was also used as a polarisation calibrator obtaining
the polarisation degree $m$ and the polarisation angle $\chi$ in agreement with the values in the literature 
\citep{Tabara} within errors $<$1 per cent and $<$5 deg, respectively. The unpolarised planetary nebula 
NGC\,7027 was also observed to have an estimation of the instrumental polarisation at each frequency.

The measurement of flux densities from the single-dish cross-scans was done by fitting Gaussians to the signal from all 
the polarimeter outputs (Stokes I, Q and U) and identifying the Gaussian amplitudes with the flux densities $S_{I}$, 
$S_{Q}$ and $S_{U}$. For all sources with significant $S_{Q}$ and $S_{U}$ contributions, the polarised flux density $S_{P}$, 
the degree of linear polarisation $m$ and the polarisation angle $\chi$ were computed, using the expressions
given in \cite{Klein}. 

\subsection{Very Large Array}\label{VLA}

VLA data were taken during February/March 2006 in the most extended A configuration using the receivers corresponding
to the X, U, K and Q bands, i.e., at 8.4, 15, 22 and 43 GHz, respectively. Integration times were chosen on the basis
of the predicted flux densities at these frequencies, after extrapolating from the Effelsberg frequencies with the adequate 
spectral indices. Those sources with indications of measurable polarised emission were observed more deeply and those
expected to be too faint for this purpose were observed to have at least a good signal to noise ratio in $S_{I}$. 
Integration times range from $\sim$1 minute at 8.4 GHz up to $\sim$1 hour at 22 or 43 GHz for some sources. 

At the two highest frequencies the fast-switch method was used, quickly switching between target and calibrator 
in order to have phase stability, taking care of the rapid atmospheric fluctuations. The duration of the cycles 
target-calibrator was of $\sim$ 150 to 200 seconds in the two highest frequencies.

3C\,286 was the primary flux density calibrator and 3C\,48 was used to test the goodness of the flux scale solution. 
Several phase calibrators chosen from the VLA calibrator manual\footnote{Available at http://www.vla.nrao.edu/astro/calib/manual} 
were observed during the run. Most were selected at
about 2-5 deg from their target sources, and in all cases within 10 deg. This is 
especially important to avoid the loss of coherence at the highest frequencies. Again the flux density scale is the one 
of \cite{Baars} except at 22 and 43 GHz where the Baars expressions are not valid. At these frequencies the adopted scale
is based on emission models and observations of planet Mars, and is supposed to be accurate at a level of 5-10 per cent.

\begin{table}
 \centering
 \begin{minipage}{140mm}
  \caption{Observing frequencies and beam sizes.}\label{observations}
  \begin{tabular}{c c r c}
  \hline
    Telescope & \multicolumn{1}{c}{Frequency}  &  \multicolumn{1}{c}{Bandwidth} & $\theta_{\rm{HPBW}}$  \\ 
              &    (GHz)   &   (MHz)       &   (arcsec)          \\ 
\hline  
Effelsberg & 2.65 &   350  & 265  \\
Effelsberg & 4.85 &   500  & 145  \\
Effelsberg & 8.35 &  1200  &  80  \\
Effelsberg & 10.5 &   300  &  65  \\
\hline
VLA(A)     & 8.45 &   700  & 0.24 \\
VLA(A)     & 14.5 &   700  & 0.14 \\
VLA(A)     & 22.5 &  2000  & 0.08 \\
VLA(A)     & 43.5 & 10000  & 0.05 \\
\hline
\end{tabular}
\end{minipage}
\end{table}

The 31DEC06 version of the AIPS package was used to flag, calibrate, clean and image the VLA data. The standard recipe was 
followed in all frequencies. However, at 22 and 43 GHz opacity and gain curve corrections were applied as a previous step to 
the standard calibration, and the calibration table was gridded in intervals of 3 seconds to get rid of rapid atmospheric
fluctuations. The appropriate clean component models were used for 3C\,286 and 3C\,48 at these two frequencies. The peak flux 
densities were measured by fitting 2-D Gaussians to the source profiles. The integrated flux densities were computed 
using the BLSUM task, adding the signal in small boxes containing each source. The same boxes were used to 
measure the Stokes parameters S$_{Q}$ and S$_{U}$. The local noise of the maps was estimated from a circular region around each 
individual source. Peak and integrated flux densities have essentially the same values because most sources are unresolved,
as will be discussed below. 

The polarisation calibration of the VLA data was determined using the AIPS tasks PCAL and RLDIF. The first solves for the 
feed parameters in each antenna, using repeated observations of a strong unresolved source (in our case several snapshots of 
3C\,286 were taken) covering a range of different parallactic angles. RLDIF determines and subtracts any phase difference 
between the right and left polarisation systems. The residual instrumental polarisation at the VLA frequencies was estimated 
measuring the linear polarisation of 3C\,84 after having applied the polarisation calibration. \cite{Aller} report that this 
source shows a percentage of linear polarisation of 0.12 $\pm$ 0.01 at 14.5 GHz, and for our purposes it can be considered 
unpolarised.

\subsection{Error determination}

We consider three main contributions to the flux density error as in \cite{Klein}. These are (i) the calibration error 
$\Delta S_{cal}$ (in percentage) which is estimated by the dispersion of the different observations of the flux density 
calibrators; (ii) the error introduced by noise, $\Delta S_{n,i}$, which is estimated from the local noise around the 
source; and (iii) the confusion error $\Delta S_{conf,i}$ due to background sources within the beam area. This last term 
can be neglected in interferometric data where the synthesised beam has small dimensions.

Thus the expressions for the total uncertainty of Stokes parameters are Equations \ref{err} and \ref{err2} for 
Effelsberg and the VLA, respectively:

\begin{equation}\label{err}
\Delta S_{i}= \sqrt{(S_{i} \cdot \Delta S_{cal})^{2} + \Delta S_{n,i}^{2} + \Delta S_{conf,i}^{2}} \hspace{0.5cm} i={\rm I,Q,U}
\end{equation}

\begin{equation}\label{err2}
\Delta S_{i}= \sqrt{(S_{i} \cdot \Delta S_{cal})^{2} + S_{n,i}^{2} \cdot \frac{A_{src}}{A_{beam}}} \hspace{0.5cm} i={\rm I,Q,U}
\end{equation}

where $A_{src}$ is the area of the box used to measure the source flux density, and $A_{beam}$ is the beam area. The 
expressions for the uncertainties of $S_{P}$, $m$ and $\chi$ can be taken from \cite{Klein}.

\begin{table}
 \centering
 \begin{minipage}{70mm}
  \caption{Values used to calculate the errors of the Stokes parameters and obtained instrumental polarisations.}\label{table_errors}
  \begin{tabular}{@{}ccccccc@{}}
\hline
Telescope  &       Frequency  &  \multicolumn{2}{c}{$\Delta S_{conf}$} &    $\Delta S_{cal}$ & $P_{instr}$ \\
           &            (GHz) &     (I)           &     (Q,U)          & \multicolumn{2}{c}{(per cent)}      \\
\hline
Effelsberg &  2.65   & 1.5   & 0.5  & 0.8 & 0.7 \\   
Effelsberg &  4.85   & 0.45  & 0.15 & 0.9 & 0.7 \\
Effelsberg &  8.35   & 0.23  &  $-$ & 1.0 & 0.6 \\
Effelsberg & 10.5    & 0.08  &  $-$ & 1.9 & 0.8 \\
\hline
VLA        &  8.45   & $-$   &  $-$ & 1.9 & $<$0.1 \\
VLA        & 15.0    & $-$   &  $-$ & 2.9 &    0.6 \\
VLA        & 22.5    & $-$   &  $-$ & 1.8 &    0.7 \\
VLA        & 43.5    & $-$   &  $-$ & 5.3 &    2.1 \\
\hline
\end{tabular}
\end{minipage}
\end{table}

Table \ref{table_errors} shows the coefficients used for the calculation of errors $\Delta S_{i}$. The confusion limits at 2.7, 4.8 and 
10.5 GHz have been extracted from \cite{Klein} and that at 8.35 GHz (not available in \citealt{Klein}) was obtained 
from our own observations inspecting the 
behaviour of the signal rms at long integration times. This value is consistent with the interpolation over frequency from 
the two adjacent bands (at 4.85 and 10.5 GHz). 
The last column in Table \ref{table_errors} lists the instrumental polarisation for the different frequencies and
telescopes.

\section{Results}\label{sec4}

\subsection{Radio flux densities}

Tables \ref{listfluxeslit} and \ref{listfluxes} show the measured flux densities and errors (in mJy) between 74 MHz
and 43 GHz for the RBQ sample. At frequencies below 2.6 GHz there are available flux densities from the literature 
and these are listed in Table \ref{listfluxeslit}. The flux densities at 325 MHz were obtained from the Westerbork
Northern Sky Survey (WENSS; \citealt{Rengelink}). In addition, 3-$\sigma$ upper limits and a detection from the 
Texas survey \citep{Douglas} at 365 MHz are given for those sources not covered by the WENSS survey. At 74 MHz none 
of the sources is detected in the VLA Low-Frequency Sky Survey (VLSS; \citealt{Cohen}), but 3-$\sigma$ upper limits 
have been computed measuring the local noise around the source position in the available images. Flux densities at 
1.4 GHz from FIRST and NVSS \citep{Condon} can be also found in Table \ref{listfluxeslit}. 

The BAL QSO 1624+37 was observed in this campaign at 2.65 GHz with Effelsberg and at 15 and 43 GHz with the VLA. 
All the flux densities in the remaining frequencies were presented by \cite{Benn}.

\begin{table}
 \begin{minipage}{80mm}
  \caption{RBQ sample of 15 radio-loud BAL QSOs: Flux densities (in mJy), errors or 3-$\sigma$ upper limits from
  74 MHz to 1.4 GHz obtained from the literature.}\label{listfluxeslit}
\begin{tabular}{lcrrrr}
\hline
\multicolumn{1}{c}{ID}        &  
\multicolumn{1}{c}{$S_{0.074}^{\rm{VLSS}}$}  & 
\multicolumn{1}{c}{$S_{0.365}^{\dagger}$} &  
\multicolumn{1}{c}{$S_{1.4}^{\rm{FIRST}}$} &  
\multicolumn{1}{c}{$S_{1.4}^{\rm{NVSS}}$} \\
\hline
0039$-$00 & $<$475  &  \multicolumn{1}{c}{$<$80}&  21.2 & 19.5 $\pm$ 0.7 \\
0135$-$02 & $<$350  &  \multicolumn{1}{c}{$<$80}&  22.4 & 22.8 $\pm$ 1.1 \\
0256$-$01 & $<$385  &  \multicolumn{1}{c}{$<$80}&  27.6 & 22.3 $\pm$ 0.8 \\
0728+40   & $<$300  &  15 $\pm$ 3.3$^{\dagger}$ &  17.0 & 17.2 $\pm$ 0.6 \\
0837+36   & $<$290  & 184 $\pm$ 3.4$^{\dagger}$ &  25.5 & $-$            \\
0957+23   & $<$260  &  \multicolumn{1}{c}{$<$80}& 136.1 &136.9 $\pm$ 4.1 \\
1053$-$00 & $<$500  &  \multicolumn{1}{c}{$<$80}&  24.7 & 22.7 $\pm$ 1.1 \\
1159+01   & $<$320  & 887 $\pm$ 30$^{~~}$       & 268.5 &275.6 $\pm$ 8.3 \\
1213+01   & $<$340  &  \multicolumn{1}{c}{$<$80}&  21.5 & 27.5 $\pm$ 7.9 \\
1228$-$01 & $<$340  &  \multicolumn{1}{c}{$<$80}&  29.4 & 29.1 $\pm$ 1.0 \\
1312+23   & $<$300  &  \multicolumn{1}{c}{$<$80}&  43.3 & 46.5 $\pm$ 1.4 \\
1413+42   & $<$330  &  16 $\pm$ 3.2$^{\dagger}$ &  17.8 & 16.8 $\pm$ 0.6 \\
1603+30   & $<$225  &  33 $\pm$ 4.4$^{\dagger}$ &  53.7 & 54.1 $\pm$ 1.7 \\
1624+37   & $<$355  &  59 $\pm$ 4.0$^{\dagger}$ &  56.1 & 55.6 $\pm$ 1.7 \\
1625+48   & $<$290  &  26 $\pm$ 3.7$^{\dagger}$ &  25.3 & 26.0 $\pm$ 0.9 \\
\hline
\end{tabular}
The symbol $\dagger$ indicate data from WENSS at 325 MHz, the rest are from the Texas Survey 
at 365 MHz.
\end{minipage}
\end{table}

\begin{table*}
 \begin{minipage}{180mm}
  \caption{RBQ sample of 15 radio-loud BAL QSOs: Flux densities (in mJy) from 2.6 to 43 GHz obtained in this work. 
  Superscripts on the errors indicate the run number of that observation, according to the key 
  in Table \ref{table_runs}. The exception is 1624+37 (with superscript 0), for which several flux densities were 
  presented in \protect\cite{Benn}}\label{listfluxes}
\begin{tabular}{lrrrrrrrrrrrrrrrrrrr}
\hline
\multicolumn{1}{c}{ID}        &  
\multicolumn{1}{c}{$S_{2.6}$} & 
\multicolumn{1}{c}{$S_{4.8}$} & 
\multicolumn{1}{c}{$S_{8.3}$} & 
\multicolumn{1}{c}{$S_{8.4}$} & 
\multicolumn{1}{c}{$S_{10.5}$}& 
\multicolumn{1}{c}{$S_{15}$}  & 
\multicolumn{1}{c}{$S_{22}$}  &  
\multicolumn{1}{c}{$S_{43}$}\\
\hline
0039$-$00 &\multicolumn{1}{c}{$-$}&  19.3 $\pm$ 0.6$^{1}$ &  12.0 $\pm$ 0.4$^{1}$ & 12.3 $\pm$ 0.1$^{5}$ &   9.7 $\pm$ 0.4$^{1}$ &  6.2 $\pm$ 1.1$^{5}$ &  2.2 $\pm$ 1.2$^{5}$ & 1.1 $\pm$ 0.1$^{5}$ \\
0135$-$02 &\multicolumn{1}{c}{$-$}&  27.2 $\pm$ 0.6$^{1}$ &  31.9 $\pm$ 0.8$^{1}$ & 30.8 $\pm$ 0.4$^{5}$ &  29.5 $\pm$ 0.7$^{3}$ & 22.7 $\pm$ 0.8$^{5}$ & 14.7 $\pm$ 0.2$^{5}$ & 2.3 $\pm$ 0.2$^{5}$ \\
0256$-$01 &\multicolumn{1}{c}{$-$}&  12.0 $\pm$ 0.5$^{1}$ &   7.2 $\pm$ 0.3$^{1}$ &  7.2 $\pm$ 0.4$^{5}$ &   5.9 $\pm$ 0.4$^{1}$ &  4.3 $\pm$ 1.5$^{5}$ &  2.2 $\pm$ 0.3$^{5}$ & 1.4 $\pm$ 0.1$^{5}$ \\
0728+40   &   6.7 $\pm$ 1.5$^{4}$ &   6.4 $\pm$ 0.7$^{1}$ &   3.1 $\pm$ 0.2$^{1}$ &  2.5 $\pm$ 0.3$^{5}$ &   1.6 $\pm$ 0.3$^{1}$ &  1.9 $\pm$ 0.3$^{5}$ & 0.83 $\pm$ 0.06$^{5}$&\multicolumn{1}{c}{$-$}\\
0837+36   &  50.4 $\pm$ 1.6$^{4}$ &  30.2 $\pm$ 2.3$^{2}$ &  12.1 $\pm$ 0.8$^{1}$ & 12.5 $\pm$ 0.1$^{6}$ &  13.0 $\pm$ 0.4$^{2}$ &  3.3 $\pm$ 1.3$^{6}$ &  2.2 $\pm$ 0.5$^{6}$ & 1.3 $\pm$ 0.2$^{6}$ \\
0957+23   &  93.7 $\pm$ 2.4$^{4}$ &  78.7 $\pm$ 1.1$^{1}$ &  48.5 $\pm$ 1.2$^{1}$ & 53.2 $\pm$ 0.5$^{5}$ &  39.8 $\pm$ 1.2$^{1}$ & 31.4 $\pm$ 0.7$^{5}$ & 22.4 $\pm$ 0.3$^{5}$ & 9.8 $\pm$ 0.9$^{5}$ \\
1053$-$00 &  13.6 $\pm$ 1.6$^{4}$ &  15.7 $\pm$ 0.6$^{1}$ &  11.8 $\pm$ 0.4$^{1}$ & 13.1 $\pm$ 0.1$^{5}$ &  12.3 $\pm$ 0.6$^{1}$ &  8.2 $\pm$ 1.6$^{5}$ &  5.9 $\pm$ 0.7$^{5}$ & 2.3 $\pm$ 0.7$^{5}$ \\
1159+01   & 133.9 $\pm$ 2.0$^{4}$ & 137.8 $\pm$ 1.7$^{1}$ & 158.0 $\pm$ 2.0$^{2}$ &160.8 $\pm$ 1.2$^{5}$ & 150.6 $\pm$ 3.7$^{2}$ &123.8 $\pm$ 1.6$^{5}$ &105.5 $\pm$ 1.1$^{5}$ &74.6 $\pm$ 2.1$^{5}$ \\
1213+01   &  23.0 $\pm$ 1.7$^{4}$ &  15.2 $\pm$ 0.5$^{1}$ &   9.7 $\pm$ 0.3$^{1}$ & 11.6 $\pm$ 0.1$^{5}$ &   8.2 $\pm$ 0.5$^{1}$ &  5.6 $\pm$ 1.4$^{5}$ &  3.6 $\pm$ 0.5$^{5}$ & 1.4 $\pm$ 0.2$^{5}$ \\
1228$-$01 &  19.4 $\pm$ 1.6$^{4}$ &  18.6 $\pm$ 0.5$^{1}$ &  15.5 $\pm$ 0.4$^{1}$ & 16.9 $\pm$ 0.2$^{5}$ &  13.5 $\pm$ 0.7$^{1}$ & 11.5 $\pm$ 1.6$^{5}$ &  8.1 $\pm$ 0.2$^{5}$ & 2.5 $\pm$ 0.7$^{5}$ \\
1312+23   &  28.9 $\pm$ 1.6$^{4}$ &  25.7 $\pm$ 0.6$^{1}$ &  19.8 $\pm$ 0.5$^{1}$ & 19.4 $\pm$ 0.2$^{5}$ &  16.1 $\pm$ 0.6$^{1}$ & 10.5 $\pm$ 0.3$^{5}$ &  7.7 $\pm$ 0.1$^{6}$ & 5.4 $\pm$ 0.4$^{6}$ \\
1413+42   &   7.6 $\pm$ 2.9$^{4}$ &   8.8 $\pm$ 0.7$^{2}$ &  13.4 $\pm$ 0.3$^{2}$ & 13.4 $\pm$ 0.1$^{6}$ &  12.7 $\pm$ 0.4$^{2}$ &  9.5 $\pm$ 1.5$^{6}$ &  7.0 $\pm$ 0.2$^{6}$ & 2.2 $\pm$ 0.4$^{6}$ \\
1603+30   &  22.8 $\pm$ 1.7$^{4}$ &  26.1 $\pm$ 0.7$^{1}$ &  19.1 $\pm$ 0.5$^{1}$ & 22.1 $\pm$ 0.3$^{5}$ &  16.8 $\pm$ 0.6$^{1}$ & 12.5 $\pm$ 1.7$^{5}$ &  8.3 $\pm$ 0.1$^{5}$ & 1.9 $\pm$ 0.5$^{5}$ \\
1624+37   &  33.5 $\pm$ 1.6$^{4}$ &  23.3 $\pm$ 1.1$^{0}$ &\multicolumn{1}{c}{$-$}& 15.0 $\pm$ 0.1$^{0}$ &  10.5 $\pm$ 0.8$^{0}$ &  9.6 $\pm$ 0.6$^{5}$ & 5.41 $\pm$ 0.02$^{0}$& 2.1 $\pm$ 0.3$^{5}$ \\
1625+48   &  17.6 $\pm$ 1.5$^{4}$ &   9.4 $\pm$ 0.7$^{2}$ &   8.0 $\pm$ 0.2$^{2}$ &  7.0 $\pm$ 0.2$^{5}$ &   6.3 $\pm$ 0.2$^{2}$ &  5.8 $\pm$ 1.3$^{5}$ &  1.9 $\pm$ 0.3$^{5}$ & 1.7 $\pm$ 0.2$^{5}$ \\
\hline
\end{tabular}
\end{minipage}
\end{table*}

\subsection{Morphologies}\label{morfo}

\begin{figure}
  \includegraphics[width=84mm]{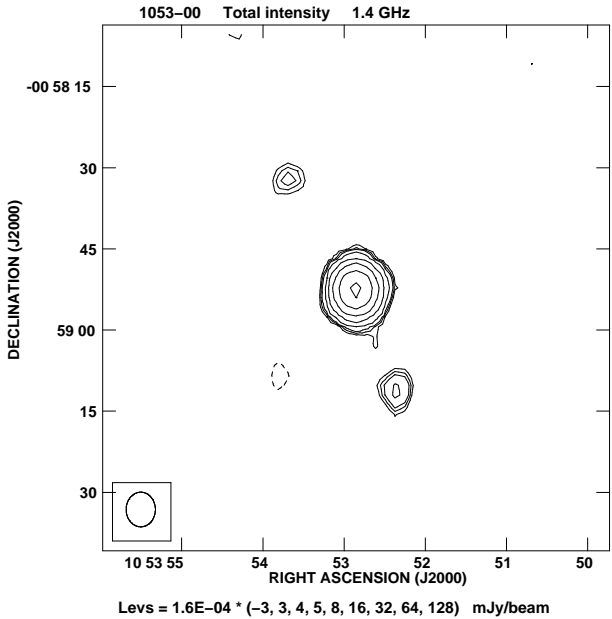}
  \caption{Map of BAL QSO 1053$-$00 from FIRST at 1.4 GHz. The synthesized beam-size is shown in the lower-left 
  corner.}\label{map1053}
\end{figure}

The FIRST maps at 1.4 GHz show a compact morphology for all 15 BAL QSOs in the RBQ sample, at a resolution of 
$\sim$5$\arcsec$. To quantify this we have computed the morphological parameter $\Theta$, based on the FIRST 
integrated and peak flux densities, $\Theta$ = log(S$_{\rm{int}}$/S$_{\rm{peak}}$), and for all 15 sources 
$\Theta <$0.03. All sources, with the exception of 1053$-$00, appear essentially point-like at this resolution.
The FIRST map of 1053$-$00 is shown in Figure \ref{map1053} where a point-like core and two faint components 
(5- and 8-$\sigma$ detections) can be seen. The faint components seem to be real and this source has been 
included in the catalogue of double-lobed radio quasars from SDSS \citep{DeVries}. This is thus an additional 
example of the rare class of BAL QSOs showing double-lobed morphology, and it is not included in the list of 8 SDSS 
BAL QSOs of this type compiled by \cite{Gregg2}. 

Our VLA maps at higher frequencies confirm the compactness of the 15 BAL QSOs, having all of them point-like 
structure at 8.4 and 15.0 GHz. The faint components of 1053$-$00 cannot be seen at these frequencies, probably 
due to a steep spectral index in these two regions.

\begin{figure}
  \includegraphics[width=84mm]{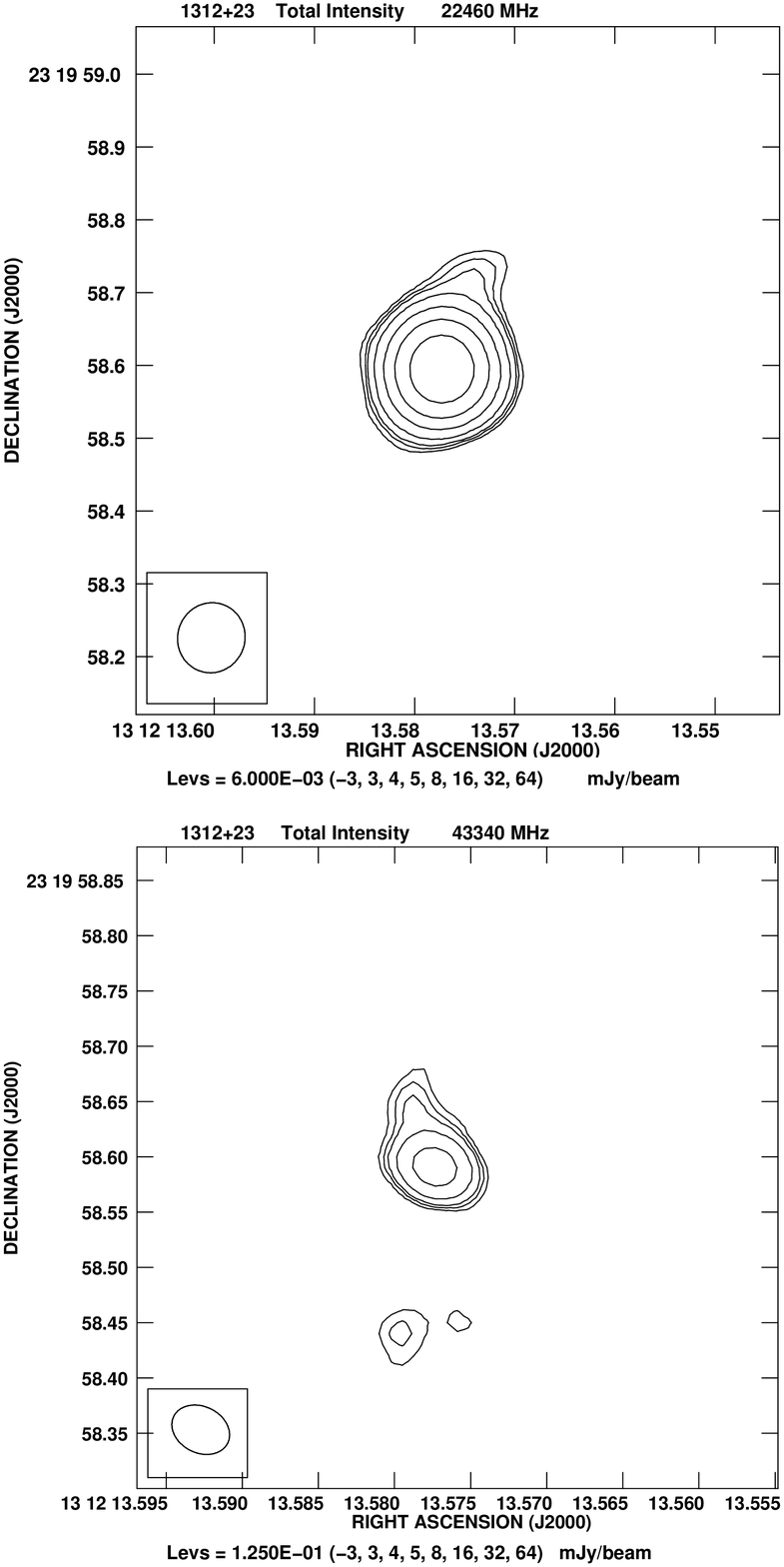}
  \caption{VLA map of BAL QSO 1312+23 at 22 GHz (top) and at 43 GHz (bottom). The synthesized beam-size is
  shown in the lower-left corner of each map.}\label{map}
\end{figure}

The combination of the 43-GHz band with the A configuration provides the highest resolution available with the VLA 
($\theta_{HPBW} \sim 50$ mas). However, as can be seen in Table \ref{listfluxes} most sources are very faint at 
43 GHz, i.e., close to the 1 mJy level. For those the self-calibration process is not appropriate and the way 
to recover the flux densities was to assume an initial point-source model before the first mapping. 
This procedure is to some extent well justified since at lower frequencies the morphology is still point-like, 
although the precise astrometric and morphological information about the sources at 43 GHz is lost.

Self-calibration at high frequency was only possible for the brightest sources: 0957+23, 1159+01 and 1312+23. 
The first two sources appear point-like in the 43-GHz maps. Figure \ref{map} shows the naturally weighted maps
of 1312+23 at 43 and 22 GHz, revealing an extension towards the north coming from the central unresolved component. 
Although the feature, possibly a jet, is detected at both frequencies, its significance is higher at 43 GHz.

The small feature in the 43-GHz map at $\sim$120 mas south from the central source is probably not real because it
is located in a region affected by a stripe. It must be noticed that at this high frequency this flux density level is 
probably close to the limit where signal coherence allows accurate mapping.
For the remaining 12 sources the mapping at 43 GHz was not possible and an upper limit on their sizes was obtained
using the observations at 22 GHz ($\theta_{HPBW} \sim 80$ mas), where they are bright enough for 
mapping. For those, the task IMFIT was used to fit an elliptical Gaussian component and a zero level within a small 
box containing the radio source, yielding the major and minor axes of the fitted Gaussian, position angle 
and the formal beam-deconvolved dimensions. The uncertainty in the dimensions was estimated following Equation 1 
of \cite{White}. Eight of the twelve sources are clearly more extended than the beam size and their fitted
parameters are given in Table \ref{dec_sizes}. Also in this table are shown the dimensions of 1312+23 and 1624+37,
as measured from the map in Figure \ref{map} and from VLBA observations to be published in a subsequent paper, 
respectively. The remaining three sources (0039$-$00, 0256$-$01, 1053$-$00) plus 0957+23 and 1159+01 can be
considered strictly unresolved at high frequencies.

\begin{table}
 \centering
 \begin{minipage}{85mm}
  \caption{Angular dimensions for the extended sources.}\label{dec_sizes}
  \begin{tabular}{lcrrrl}
  \hline
Source & Freq.  & $\theta_{maj}$ & $\theta_{min}$ &  PA & $\theta_{maj}^{(dec)}$\\ 
       & (GHz)  &  (mas)      &      (mas)   &  (deg)   &      (mas)            \\ 
\hline
0135$-$02 & 22  & 112.4 & 78.7 & $-$12 & 11 $\pm$ 4\\
0728+40   & 22  &  85.5 & 73.0 &    26 & 26 $\pm$ 6\\
0837+36   & 22  &  94.0 & 85.9 &    34 & 19 $\pm$ 10\\
1213+01   & 22  & 111.7 & 75.5 & $-$19 & 27 $\pm$ 11\\
1228$-$01 & 22  &  95.3 & 76.3 & $-$30 & 19 $\pm$ 4\\
1312+23   & 43  &  54.5 & 41.8 &    64 & 69 $\pm$ 12$^{\dagger}$\\
1413+42   & 22  &  92.0 & 69.5 & $-$60 &$~$9 $\pm$ 3\\
1603+30   & 22  &  95.8 & 71.0 & $-$71 & 15 $\pm$ 3\\
1624+37   & 4.8 & $-$  & $-$  &    $-$& 36 $\pm$ 5$^{\dagger\dagger}$\\ 
1625+48   & 22  & 157.0 & 71.8 &    76 & 63 $\pm$ 20\\
\hline
\end{tabular}
\end{minipage}
$\dagger$ Not deconvolved size, but measured from the extended structure in Figure \ref{map}.
$\dagger \dagger$ Measured from VLBA map at 4.8 GHz to be presented in Montenegro-Montes et al. 
(in prep.)
\end{table}

VLBI observations of 0957+23 and 1312+23 have been presented by \cite{Jiang} using the European VLBI Network at 
1.6 GHz. With a restoring beam of only 18.5 $\times $ 6.86 mas, 0957+23 appears as a single point-like component, 
while 1312+23 is resolved. This last source shows a symmetric morphology with a central core and 
two components in the northern and southern directions with a total extension of $\sim$150 mas (i.e., about 
1 kpc). Although an extended morphology is also present in our 22-GHz and 43-GHz maps (Fig. \ref{map}), only
the northern part is detected.

Summarising, our VLA maps show very compact morphologies for all 15 sources at all frequencies, at a maximum
resolution of 80 mas. The exceptions are 1312+23 which shows some elongation at 22 and 43 GHz, and 1053$-$00 which 
is extended at 1.4 GHz showing two faint lobes not detected at higher frequencies.

\subsection{Variability}

We have checked for variability at 1.4 GHz comparing the flux densities from FIRST and NVSS epochs in those 14
sources with NVSS measurement. The same has been done at 8.4 GHz for those 5 sources in common with the list 
of \cite{Becker1}. They observed these sources with the VLA in A and D configuration 
(resolutions of 0.8 and 9 arcsec respectively) and all 5 sources presented point-like structure at both resolutions. 
As mentioned before this is again confirmed by our observations at 8.4 GHz in A configuration. This means that flux 
densities of both measurements can be directly compared. The same is true at 1.4 GHz because all sources except 
1053$-$00 are point-like at the resolution of FIRST (B configuration) and the resolution of NVSS is poorer 
(VLA D and CnB configurations). Although 1053$-$00 is resolved, it has a compact nucleus and the two extensions 
are not expected to contribute much to the total flux density (Section \ref{morfo}).

We measure the flux density variations with the parameter $Var_{\Delta S}$, as defined e.g. by \cite{Torniainen05}:
\begin{equation}\label{sigma_VR}
Var_{\Delta S} = \frac{S_{max}-S_{min}}{S_{min}}
\end{equation}
As an estimate of the significance of the source variability the $\sigma_{var}$ parameter defined by \cite{Zhou} 
is used, adopting the \emph{integrated} flux densities $S_{1}$ and $S_{2}$ at the two epochs:
\begin{equation}\label{sigma_VR}
\sigma_{Var} = \frac{|S_{2}-S_{1}|}{\sqrt{\sigma_{2}^{2}+\sigma_{1}^{2}}}
\end{equation}
where $\sigma_{2}$ and $\sigma_{1}$ are the uncertainties in the integrated flux densities. Table \ref{variability} 
shows flux densities, $Var_{\Delta S}$ and $\sigma_{Var}$ at 1.4 and 8.4 GHz, for those sources exhibiting 
$\sigma_{Var}>$ 3. For the flux densities of \cite{Becker1}, whose errors are not given in their work, a 5 per 
cent error is assumed.

A low percentage of the sample (only 2 out of 14 sources) shows variability at 1.4 GHz, both sources varying about 
20 per cent and with a significance just above 3 standard deviations. At 8.4 GHz, 3 out of 5 sources are variable 
at this level and 1312+23 constitutes the most extreme case, showing variations of about 50 per cent with 
significance above 10.

\begin{table}
 \centering
 \begin{minipage}{85mm}
  \caption{Sources showing significant flux density variability. At 1.4 GHz, $S_{1}$ comes from NVSS and $S_{2}$ from
  FIRST. At 8.4 GHz, $S_{1}$ comes from \protect\cite{Becker1}, $S_{2}$ from this work (Table \ref{listfluxes})}\label{variability}
  \begin{tabular}{@{}lcccrr@{}}
\hline 
\it Source & Freq &   $S_{1}$    &  $S_{2}$         & $Var_{\Delta S}$ & $\sigma_{Var}$\\
\it        & (GHz)&      (mJy)     &    (mJy)         &                  &               \\
\hline
0256$-$01  & 1.4  &22.3$\pm$0.8 D& 27.5 B&   0.23  &3.2\\
1213+01    & 1.4  &27.5$\pm$0.9 D& 22.9 B&$-$0.20  &3.2\\
\hline 
1312+23    & 8.4  &    12.6 A    & 19.4$\pm$0.2 A &0.54& 10.3\\
1413+42    & 8.4  &    11.3 D    & 13.4$\pm$0.1 A &0.19&  3.7\\
1603+30    & 8.4  &    18.1 A    & 22.1$\pm$0.3 A &0.22&  4.2\\
\hline 
\end{tabular}
\end{minipage}
\end{table}

Flux density variations have recently been used to constrain the orientation of several BAL QSOs. \cite{Zhou} and 
later \cite{Ghosh} found a few examples of BAL QSOs with variability as high as 40 per cent at 1.4 GHz between 
the NVSS and FIRST, i.e., in periods of about 1-5 years. These variations were high enough to make the brightness 
temperature $T_{B}$ exceed the $10^{12}$ K limit associated to the Compton catastrophe. This was interpreted as 
a sign of beaming and the jet was supposed to be oriented at a few degrees from the line of sight. 

The same procedure can be applied to 0256$-$01 and 1213+01, accepting the 3-$\sigma$ significance as real.
Following the formalism in \cite{Ghosh} and knowing that the time differences between the NVSS 
and FIRST observations are 2.1 and 3.4 years, respectively, the observed variations imply brightness 
temperatures of $T_{b}$=3.1$\cdot$10$^{13}$ K for 0256$-$01 and 1.3$\cdot$10$^{13}$ K for 1213+01. 
These translate into upper limits for the jet orientation angle with respect to the line of sight of $\sim$20 
and 25 deg, respectively.
One concern about this approach is that it is based on variability at a single frequency and from only two
measurements. Moreover, the flux densities we obtained for 0256-01 at 8.4 GHz, with Effelsberg and almost one
year later with the VLA, show a good match, i.e., 7.2$\pm$0.3 and 7.2$\pm$0.4 mJy respectively, in contradiction
with the expected stronger variability at high frequencies.

\subsection{Shape of the radio spectra}\label{subsec_sp}

\subsubsection{Sample of radio BAL QSOs}\label{subsec_sp_1}

\begin{figure*}
  \includegraphics[width=180mm]{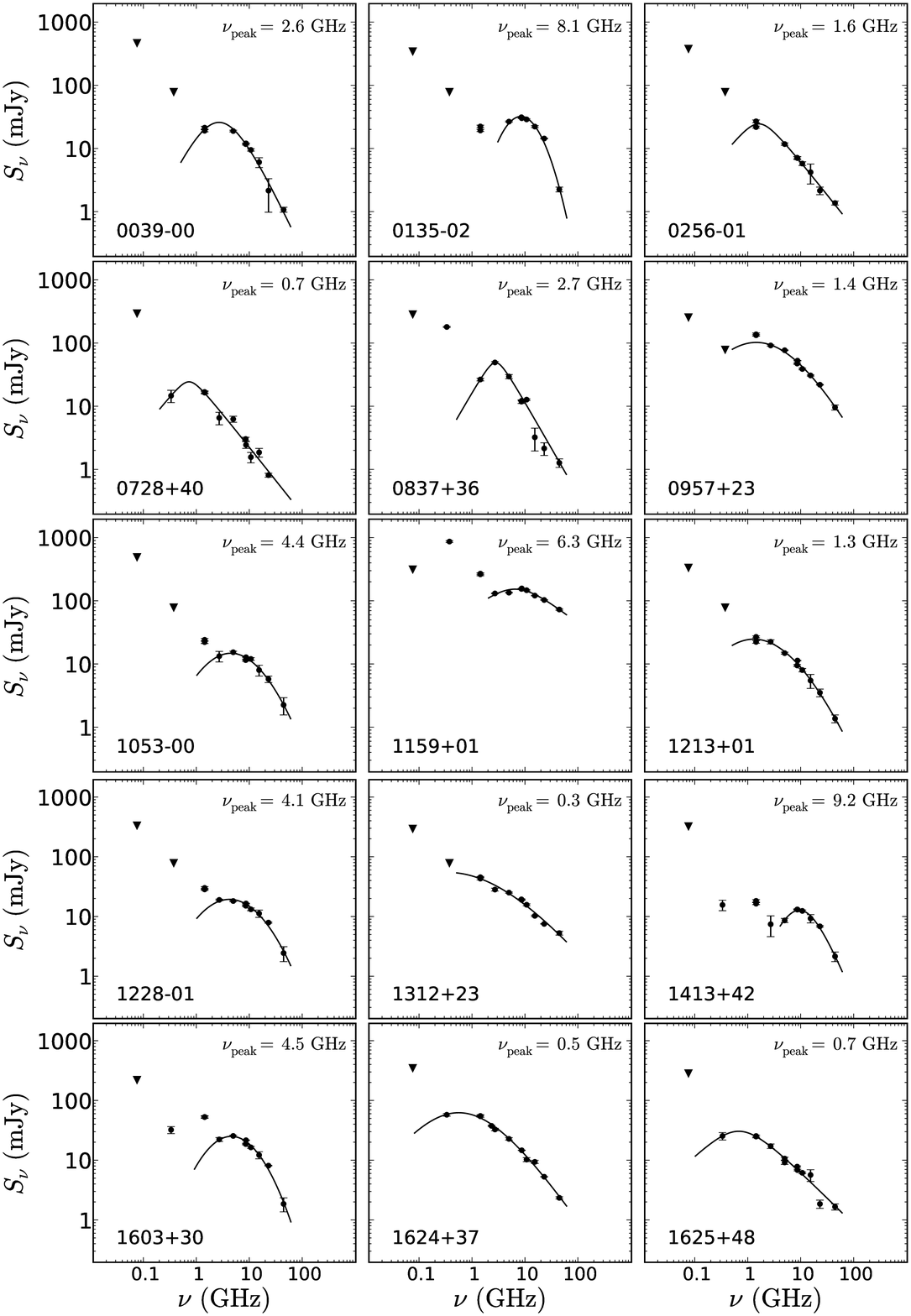}
  \caption{Radio spectra of the 15 BAL QSOs in the RBQ sample. Errors are shown if larger than the symbol size. 
  Triangles mean 3-$\sigma$ upper limits. Solid lines are fits to the analytical function give in Section 
  \ref{subsec_sp_1}. The peak frequencies given at the top-right of each pannel were obtained from the fit. 
  Some sources like 1159+01, 1413+42 and 1603+30 seem to have a second component at low 
  frequencies.}\label{synage_fits}
\end{figure*}

Figure \ref{synage_fits} show the spectra of the 15 objects. Many of them have a relatively steep spectrum at 
high frequency. About 75 per cent of the sample show flattening of the spectrum at low frequencies (typically 
below 1-5 GHz) which could indicate synchrotron self-absorption. However, in some cases some variability cannot 
be excluded, since the 1.4-GHz measurements were taken some years before the others. About 1/3 of the sample 
show, in addition, enhanced emission at MHz frequencies which could be indications of a second component emitting 
mostly in this range. Again, variability cannot be ruled out, but any possible time variability is expected to be 
less pronounced at lower frequencies. The source 1159+01 is the one which best fits in this category with both, 
a depression of the spectrum towards low frequency and a contribution from a second component in its spectrum.

Following the approach by \cite{Dallacasa}, we have fitted each spectrum with the analytical function:

\begin{equation}
log S_{\nu} = a - \sqrt{b^{2}+(c~log(\nu) - d)^{2}}
\end{equation}
where $a$,$b$,$c$,$d$ are numerical constants. This function, presented in \cite{Dallacasa} was used by these
authors to obtain the frequency turnover, $\nu_{\rm{peak}}$, of sources with convex spectra peaking above a 
few GHz, referred to in that paper as High Frequency Peakers. The value of $\nu_{\rm{peak}}$ approximately 
marks the frequency region where the source starts to be optically thick to synchrotron radiation.

These fits have been done excluding some points in order to obtain a reduced $\chi^2$ below 0.05. This was
possible for most of the sources which present a smooth radio spectrum, specially at high frequency, but not 
for 0728+40 ($\chi^2$=0.08) which presents a large scatter. At low frequencies, as we noted before, there 
are sources for which a possible second component arises. Again in these cases we decided whether to include 
or not in the fit these (low frequency) points trying to maintain the reduced $\chi^2$ within the mentioned 
limit. An exception was done with 0957+23 because even if an acceptable $\chi^2$ was obtained when including the 
1.4 GHz flux densities, the resulting fit was not compatible with the 365 MHz upper limit. 

In our sample we find 9 sources for which the spectrum peaks at a frequency higher than 1 GHz in the 
observer's frame and there are 4 displaying a peak in the range 300-700 MHz. For the remaining three objects 
(0256$-$01, 0957+23 and 1312+23) the spectrum is slightly convex but not showing an obvious peak. However, 
power-law fits would not be compatible with the Texas upper limits and then extrapolated peaks compatible 
with both the spectrum curvature and these upper limits have been determined. The peak frequencies have
to be multiplied by (1+z) to obtain intrinsic peak-frequencies.

\begin{figure}
  \includegraphics[width=90mm]{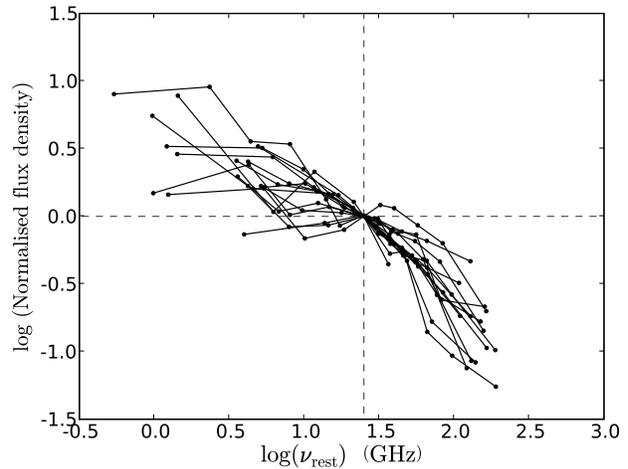}
  \caption{Rest-frame spectra of 15 BAL QSOs in the RBQ sample. They have been normalised to
  $\nu_{rest}$=25.0 GHz (log($\nu_{rest}$)=1.398)}\label{rest_SEDs}
\end{figure}

In order to understand the global trends on the spectral shape of BAL QSOs, we compare in Figure 
\ref{rest_SEDs} all the spectra bringing them first to the rest-frame in order to avoid the effect of
the redshift. The rest-frame normalisation frequency was chosen to be 25 GHz. This is a compromise to choose 
a region representative of pure synchrotron emission, where neither synchrotron self-absorption nor 
synchrotron losses might be present in the majority of sources. This frequency also makes the dispersion 
of the normalised spectra relatively small compared to other normalisation frequencies. 

We see in Figure \ref{rest_SEDs} a variety of spectral slopes around 25 GHz. There is one object in the sample 
which differ more strongly from the main trend, This is 1413+42 which has a spectrum with two components 
separated at 4 GHz (see Figure \ref{synage_fits}), and for this reason it shows in Figure \ref{rest_SEDs} an 
inverted spectrum around the normalisation frequency. Apart from this object, and assuming that variability 
effects are not very important, Figure \ref{rest_SEDs} reflects the rest-frame spectrum of a typical BAL QSO 
in the RBQ sample. It is in general convex, quite flat below 10 GHz, and becoming steeper around 25 GHz. About 
half of the objects become even steeper at frequencies higher than 50 GHz. 

The spectral shape can be quantified by means of the spectral indices. Table \ref{listalpha} provides the
turnover frequencies and representative spectral indices at different frequencies, both in the observer's 
and in the rest-frame. The spectral indices $\alpha_{0.365}^{1.4}$, $\alpha_{4.8}^{15}$ and $\alpha_{8.4}^{43}$ 
describe the low-, medium- and high-frequency regions. Moreover, the index $\alpha_{1.4}^{8.4}$ is shown for 
comparison with \cite{Becker1}. It is worth noting that the spectral indices $\alpha_{1.4}^{8.4}$ shown in 
Table \ref{listalpha} are \emph{not} based on simultaneous observations, but as we previously have shown, 
variability might not be very important at this frequency, and in any case it only affects a small number 
of sources. In the rest-frame the indices $\alpha_{6.0}^{25}$, $\alpha_{12}^{25}$, $\alpha_{25}^{50}$ and 
$\alpha_{50}^{100}$ are also given in Table \ref{listalpha}.

Looking at the rest-frame spectral indices of BAL QSOs and Figure \ref{rest_SEDs}, we see on average a 
flattening below 10 GHz, which is probably due to synchrotron self-absorption, and an overall convex shape. 
The low-frequency flattening is reflected in the median $\alpha_{6}^{25}$=$-$0.32, while the median 
$\alpha_{12}^{25}$ decreases to $-$0.57. The steepening at high frequency can be seen from the median 
$\alpha_{25}^{50}$ dropping to $-$0.92 and finally becoming steeper at higher frequencies with a median 
$\alpha_{50}^{100}$=$-$1.24. 

The radio spectral index is known to be a statistical indicator of the orientation of radio sources \citep{Orr}.
\cite{Becker1} found respectively two thirds and one third of the radio-loud BAL QSOs in their list with steep 
and flat spectral index. However, it is worth to note that they define the limit between flat and steep to be
$\alpha$=$-$0.5, and about one third of the sample has spectal index close to this value 
$-$0.6 $\le \alpha \le -$0.4. They suggest that this result is not consistent with BAL QSOs being oriented 
along a particular direction with respect to the line of sight to them. We find in our sample, which overlaps
with that of \cite{Becker1} in 5 objects, 60 per cent (9/15) of flat and 40 per cent of steep (6/15) spectrum 
sources, with 27 per cent of objects (4/15) being in the range $-$0.6 $\le \alpha \le -$0.4. These percentages
are roughly consistent within the errors, which are large because the samples are relatively small. 

\begin{table*}
 \centering
 \begin{minipage}{135mm}
  \caption{Turnover frequencies (in GHz) and spectral indices for the RBQ sample in the observer's frame
  (columns 2-6) and the rest frame (columns 7-11). The spectral indices in the observer's frame have been 
  computed from each pair of observed flux densities while the rest-frame spectral indices are based on 
  interpolation over the whole spectrum. Median values of the spectral indices are given in the last row.}\label{listalpha}
\begin{tabular}{lrrrrrrrrrr}
\hline
\multicolumn{1}{c}{}  &  
\multicolumn{5}{c}{Observer's frame}  &  
\multicolumn{5}{c}{Rest frame}  \\
\multicolumn{1}{c}{ID}  &  
\multicolumn{1}{c}{$S_{peak}$}  &  
\multicolumn{1}{c}{$\alpha_{0.365}^{1.4}$} &
\multicolumn{1}{c}{$\alpha_{1.4}^{8.4}$} &  
\multicolumn{1}{c}{$\alpha_{4.8}^{15}$} & 
\multicolumn{1}{c}{$\alpha_{8.4}^{43}$} & 
\multicolumn{1}{c}{$S_{peak}$} & 
\multicolumn{1}{c}{$\alpha_{6.0}^{25}$} &  
\multicolumn{1}{c}{$\alpha_{12}^{25}$} &
\multicolumn{1}{c}{$\alpha_{25}^{50}$} &
\multicolumn{1}{c}{$\alpha_{50}^{100}$}\\
\multicolumn{1}{c}{(1)}  &  
\multicolumn{1}{c}{(2)}  &  
\multicolumn{1}{c}{(3)}  &  
\multicolumn{1}{c}{(4)}  &  
\multicolumn{1}{c}{(5)}  &  
\multicolumn{1}{c}{(6)}  &  
\multicolumn{1}{c}{(7)}  &  
\multicolumn{1}{c}{(8)}  &  
\multicolumn{1}{c}{(9)}  &  
\multicolumn{1}{c}{(10)} &
\multicolumn{1}{c}{(11)}\\
\hline
0039$-$00  	      & 2.60 &  $-$    &  $-$0.30  &$-$1.01  &$-$1.48& 8.45 &$-$0.32  &$-$0.54  &$-$1.21  &  $-$1.88\\
0135$-$02  	      & 8.15 &  $-$    &    0.18   &$-$0.16  &$-$1.59&22.90 & +0.17   & +0.18	&$-$0.70  &  $-$2.26\\
0256$-$01  	      & 1.60 &  $-$    &  $-$0.75  &$-$0.91  &$-$1.01& 5.50 &$-$0.74  &$-$0.80  &$-$0.90  &  $-$1.28\\
0728+40    	      & 0.70 & +0.08   &  $-$1.07  &$-$1.08  &  $-$  & 1.15 &$-$0.88  &$-$0.75  &  $-$    &    $-$  \\
0837+36    	      & 2.70 & $-$1.35 &  $-$0.40  &$-$1.96  &$-$1.39&12.00 &$-$0.09  &$-$1.00  &$-$1.30  &  $-$2.16\\
0957+23    	      & 1.40 &  $-$    &  $-$0.52  &$-$0.81  &$-$1.04& 4.25 &$-$0.51  &$-$0.60  &$-$0.91  &  $-$1.09\\
1053$-$00  	      & 4.40 &  $-$    &  $-$0.35  &$-$0.58  &$-$1.07&11.15 &$-$0.14  &$-$0.30  &$-$0.94  &  $-$1.31\\
1159+01    	      & 6.30 & $-$0.81 &  $-$0.29  &$-$0.09  &$-$0.47&18.90 &$-$0.09  & +0.22	&$-$0.44  &  $-$0.48\\
1213+01    	      & 1.30 &  $-$    &  $-$0.35  &$-$0.88  &$-$1.30& 4.90 &$-$0.35  &$-$0.61  &$-$1.02  &  $-$1.20\\
1228$-$01  	      & 4.15 &  $-$    &  $-$0.31  &$-$0.43  &$-$1.17&15.10 &$-$0.29  &$-$0.12  &$-$0.55  &  $-$1.12\\
1312+23    	      & 0.35 &  $-$    &  $-$0.45  &$-$0.79  &$-$0.79& 0.85 &$-$0.43  &$-$0.58  &$-$1.01  &  $-$0.57\\
1413+42    	      & 9.20 & +0.07   &  $-$0.16  & +0.07   &$-$1.11&35.05 &$-$0.23  & +0.45	&$-$0.06  &  $-$1.04\\
1603+30    	      & 4.50 & +0.33   &  $-$0.50  &$-$0.65  &$-$1.51&13.50 &$-$0.29  &$-$0.16  &$-$0.98  &  $-$1.73\\
1624+37    	      & 0.55 & $-$0.03 &  $-$0.74  &$-$0.79  &$-$1.21& 2.40 &$-$0.71  &$-$0.64  &$-$0.99  &  $-$1.00\\
1625+48    	      & 0.65 & $-$0.02 &  $-$0.72  &$-$0.43  &$-$0.87& 2.45 &$-$0.76  &$-$0.81  &$-$0.41  &  $-$1.69\\
\hline
\multicolumn{2}{l}{Median:}  & $-$0.02 &  $-$0.40  &$-$0.79  &$-$1.14&      &$-$0.32  &$-$0.57  &$-$0.92  &  $-$1.24\\
\hline
\end{tabular}

\end{minipage}
\end{table*}

\subsubsection{Comparison with non-BAL QSOs}\label{subsec_sp_2}

To further investigate the question of the orientation of BAL QSOs we want to compare in this paper the spectral 
index distribution of quasars in a pure BAL QSO sample with that of a sample of non-BAL QSOs with the idea of 
interpreting possible differences with different orientations of both samples. In this comparison we will use the 
spectral range 1.4-8.4 GHz in the observer's frame and 6.0-25 GHz in the rest frame.

The list of non-BAL QSOs will be that of \cite{Vigotti97} presented in Section \ref{sec2}, but in all the following 
discussion we will restrict it to objects with z$>$0.5. This restriction only reduces the total number of objects 
from 124 to 119. We have computed the spectral indices in the same way as in our sample, considering pairs of
observed flux densities in the observer's frame and interpolating over the whole spectra in the rest frame.
The spectral range covered by the non-BAL QSO sample goes from 74 MHz up to 10.5 GHz, and only those sources
with z$>$1.5 will reach 25 GHz in the rest frame.  

As a BAL QSO sample we will add to our RBQ sample the list of \citep{Becker1}. They only provide flux densities
in two frequencies, i.e., 1.4 and 8.4 GHz. Thus, combining these we end up with a combined sample of 38 BAL QSOs. 
These spectral indices can be found in Table \ref{listalpha} for objects in the RBQ sample, and in Table 2 of 
\citep{Becker1} for the remaining objects. 

The Vigotti et al. sample could in principle contain both BAL QSOs and non-BAL QSOs. However, a low contamination 
of BAL QSOs is expected. \cite{Becker2} showed that in the radio-powerful regime the fraction of BAL QSOs is small. 
They used the definition of radio-loudness 
parameter $R^{*}$ given by \cite{Stocke}, i.e., the ratio of radio power at 5 GHz to optical luminosity at 
2500 \AA\, in the rest frame. In the regime of QSOs with $log(R^{*})>$2 they estimated a fraction of 4 per cent 
($<$1.5 per cent) of HiBAL (LoBAL) QSOs from the FBQS, in the interval of redshift [1.5,4.0] ([0.5,1.5]) 
where the C\,{\sc iv} (Mg\,{\sc ii}) absorbing features would be covered in optical spectra.

We have calculated $log(R^{*})$ for all quasars in Vigotti's list. For this purpose, the radio luminosity at 
5 GHz has been computed on the basis of the observed radio spectra, applying the appropriate K-correction
interpolating between each pair of observed frequencies with the adequate spectral index. The optical luminosities 
at 2500 \AA\, were computed using the POSS-I APM E magnitudes published in \cite{Vigotti97}, but corrected from 
Galactic extinction following the maps of \cite{Schlegel} and assuming an optical spectral index of 
$\alpha_{opt}=-1$ ($S_{\nu} \propto \nu^{\alpha}$), which is the value adopted by \cite{Becker2}.

\begin{figure}
  \includegraphics[width=84mm]{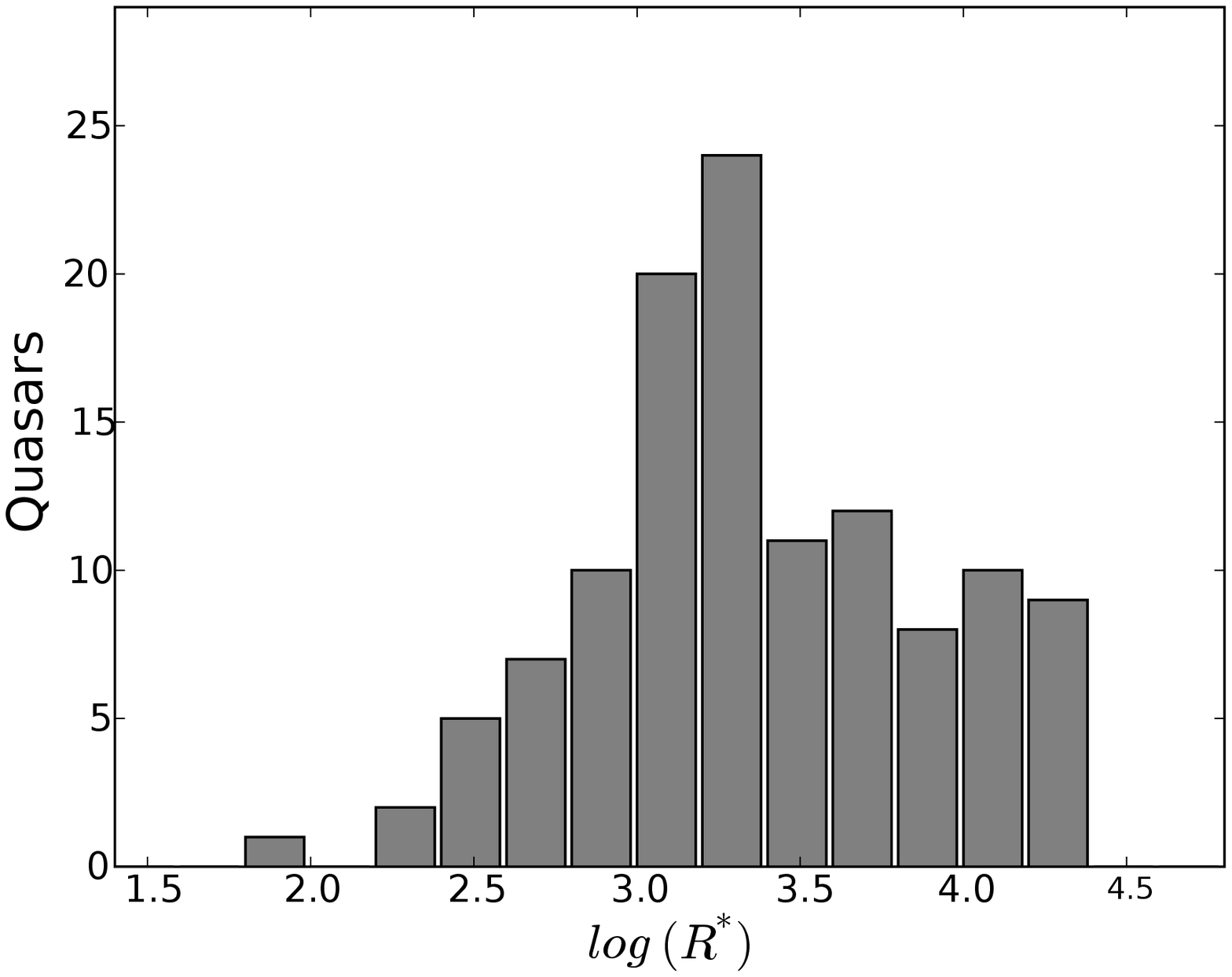}
  \caption{Histogram of radio-loudness parameter $R^{*}$ as defined by \protect\cite{Stocke} for the sample of 
  \protect\cite{Vigotti97}.}\label{hist_rloud}
\end{figure}

In Figure \ref{hist_rloud} the histogram shows how $log(R^{*})$ ranges from 1.9 to 4.6. To estimate the fraction
of BAL QSOs following the mentioned proportions, we split this sample in two groups, one with 0.5$<$z$\leq$1.5 
and the other with z$>$1.5, comprising 80 and 44 quasars respectively. About 1-2 HiBAL QSOs are expected in the 
high-z subsample and about 1 BAL QSO in the low-z subsample.

\begin{figure}
  \includegraphics[width=84mm]{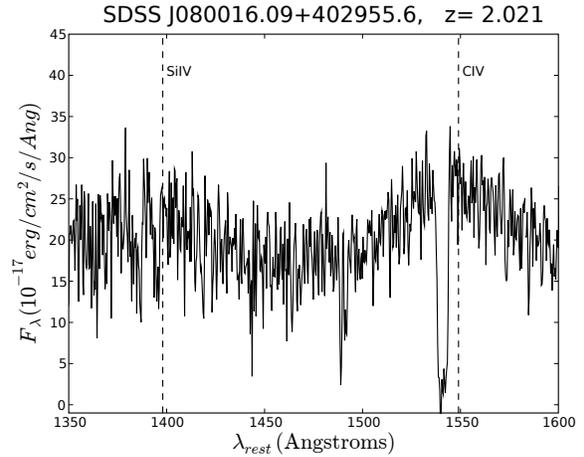}
  \caption{Spectrum of SDSS J080016.09+402955.6, showing a relatively narrow absorption in C\,{\sc IV} and 
  Si\,{\sc IV}.}\label{optical_sp}
\end{figure}

We made use of the NASA Extragalactic Database (NED) and the SDSS-DR6 database to look for the optical spectra 
of QSOs from \cite{Vigotti97} and check whether they are BAL QSOs. From 119 objects with z$<$0.5, 55 were found in
the SDSS-DR6 database. These were inspected visually to look for the presence of BAL features. All of them are 
non-BAL QSOs and only one of them, (SDSS J080016.09+402955.6, associated to the radio source B3 0756+406) shows 
some broad absorption component bluewards of the C\,{\sc iv} emission line, also present in Si\,{\sc iv}. However, 
this absorption has a width of $\sim$ 1000 km s$^{-1}$ and it is located within 3000 km s$^{-1}$ from the line
emission. It is therefore not consistent with the definition of BAL QSO but it could be better classified as a 
mini-BAL or a Narrow Absorption Line (NAL) system (see Figure \ref{optical_sp}).

We searched for the remaining 64 spectra in the NED and found 14 of them in different publications, most of them 
in \cite{Vigotti97} and \cite{Lahulla}. From these 14, only 4 could be identified as normal QSOs without BALs: 
B3 0034+393 with z= 1.94 \citep{Osmer}; B3 0219+443 with z=0.85 \citep{Djorgovski}; B3 0249+383 with z=1.12 
\citep{Henstock} and B3 2351+456 with z=2.00 \citep{Stickel}. The other 10 are difficult to classify because they 
are either faint, with low signal-to-noise ratio in the continuum (just good enough for a correct redshift 
identification), or the wavelength coverage does not include the region where the absorbing troughs could be 
present. Summing up, 59 out of 119 QSOs (roughly half of the sample) were inspected, with no one clearly showing 
the BAL phenomenon. These results justify the assumption that the list of \citep{Vigotti97} can be used as a 
representative sample of non-BAL quasars.

Table \ref{subsamples} summarises the number of elements and the basic statistics on the spectral index for the
samples under comparison. Since BAL QSOs are compact objects as \cite{Becker1} discovered and we confirm in this
work, we will also consider in our comparison a subsample of non-BAL QSOs extracted from \cite{Vigotti97} which 
includes only those compact sources with angular sizes $<$0.5 arcsec at 1.4 GHz. The angular sizes at 1.4 GHz in 
Vigotti's sample were measured from maps taken with the VLA in C and D configurations \citep{Vigotti89} and for 
the most compact sources from observations in A configuration (priv. communication).

The first four rows in Table \ref{subsamples} show that non-BAL QSOs have a median $\alpha_{1.4}^{8.4}$ lower 
than BAL QSOs ($-$0.92 and $-$0.53, respectively). The conclusion could be that BAL QSOs tend to have flatter 
spectra and are, on average, oriented at a lower angle with respect to the line of sight than normal QSOs. On 
the other hand, when the non-BAL QSOs sample is restricted to compact objects only, the median spectral index 
becomes $-$0.49, very similar to that of BAL QSOs. When the full sample of non-BAL QSOs is considered, those
extended objects have a higher contribution of emission produced in the lobes, which translates into steeper 
spectral indices for non-BAL QSOs. A similar behaviour can be found when analysing the rest-frame spectral
indices $\alpha_{12}^{25}$ and also $\alpha_{6}^{25}$ although in this last case the median values for BAL QSOs 
and compact non-BAL QSOs are more distant, i.e. $-$0.32 and $-$0.47 respectively.

\begin{table}
 \centering
 \begin{minipage}{80mm}
  \caption{Statistics on spectral indices. B00: \protect\cite{Becker1}; V97: \protect\cite{Vigotti97}; 
  V97c: sources from \protect\cite{Vigotti97} with sizes $<$5 arcsec at 1.4 GHz}\label{subsamples}
  \begin{tabular}{@{}lccrccc@{}}
\hline 
    Sample &    N &     Min & Max   & Mean     &  Median   & Std   \\
\hline
\multicolumn{7}{c}{\it Observer's frame: $\alpha$ \rm{[1.4 $-$ 8.4] GHz}}\\
\hline 
RBQ+B00      & 38 & $-$1.50 &  0.70 & $-$0.50  & $-$0.53   &  0.43 \\
RBQ          & 15 & $-$1.19 &  0.20 & $-$0.52  & $-$0.44   &  0.34 \\
V97          & 119& $-$1.47 &  0.86 & $-$0.80  & $-$0.92   &  0.43 \\
V97c         & 50 & $-$1.41 &  0.86 & $-$0.53  & $-$0.49   &  0.49 \\
\hline 
\multicolumn{7}{c}{\it Rest frame: $\alpha$ \rm{[6.0 $-$ 25] GHz}} \\
\hline 
RBQ          & 15 & $-$0.87 &  0.17 & $-$0.38  & $-$0.32   &  0.30 \\
V97          & 51 & $-$1.27 &  0.17 & $-$0.71  & $-$0.82   &  0.38 \\
V97c         & 27 & $-$1.21 &  0.17 & $-$0.53  & $-$0.47   &  0.40 \\
\hline 
\multicolumn{7}{c}{\it Rest frame: $\alpha$ \rm{[12 $-$ 25] GHz}}  \\
\hline 
RBQ          & 15 & $-$1.00 &  0.45 & $-$0.40  & $-$0.58   &  0.43 \\
V97          & 51 & $-$1.39 &  0.39 & $-$0.71  & $-$0.76   &  0.42 \\
V97c         & 27 & $-$1.31 &  0.39 & $-$0.54  & $-$0.62   &  0.44 \\
\hline 
\end{tabular}
\end{minipage}
\end{table}

\begin{figure}
  \includegraphics[width=85mm]{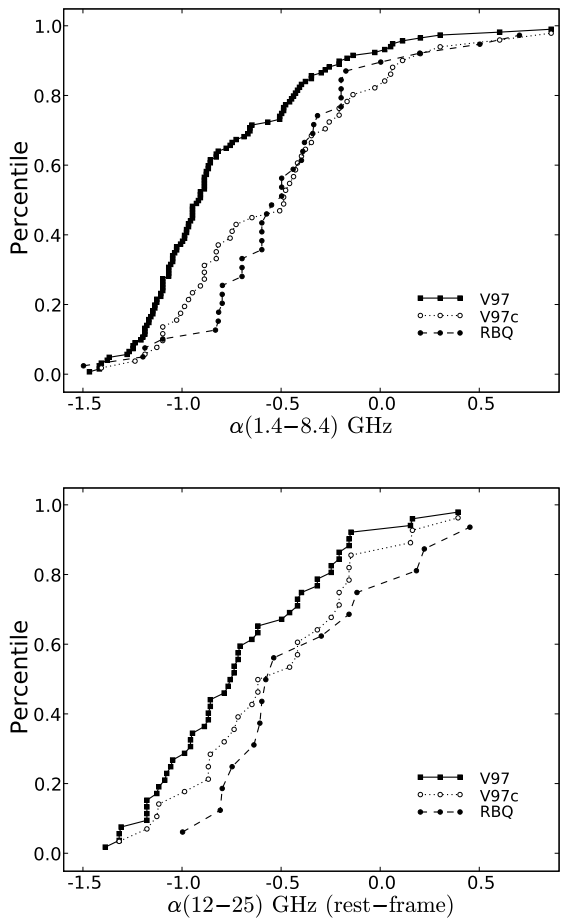}
  \caption{Percentile plot of the samples involved in the K-S tests, illustrating the shape of the spectral index
  distributions.}\label{percentile}
\end{figure}

A Kolmogorov-Smirnov (K-S) test has been done in order to compare the $\alpha_{1.4}^{8.4}$ spectral index 
distributions of these samples. The test yields a probability P$<$0.001 that both distributions are statistically
equal, confirming that they are indeed different populations. A similar K-S test restricted to only compact
sources from \cite{Vigotti97} yields P=0.204 which is consistent with both samples having a similar distribution.
The percentile plot in the upper panel of Figure \ref{percentile} illustrates the shape of the spectral index 
distribution of these three samples. It can be seen that the important contribution of extended sources from 
\cite{Vigotti97} makes the percentile curve rise more rapidly than in the other samples. This result essentially 
means that the differences in spectral indices between BAL and non-BAL QSOs arise only because BAL QSOs are compact 
sources, while normal QSOs include both compact and extended.

We will also compare the rest-frame spectral indices indices to look for possible intrinsic variations between
BAL and non-BAL QSOs, once the redshift effect has been corrected for. The most reasonable comparison is using 
$\alpha_{12}^{25}$ where the effect of synchrotron self-absorption seem less pronounced. Two K-S tests have been 
done comparing these samples. The first compares Vigotti's sources and RBQ yielding P=0.048, while the second 
compares only the compact sources in \cite{Vigotti97} and RBQ, yielding P=0.629. Again we can talk about 
statistically different populations when all QSOs are included in the comparison, but not when only spectral 
indices of compact sources are compared. The percentile plot is shown in the lower panel of Figure \ref{percentile}.

Even if the previous statistical test are consistent with similar spectral index distributions for BAL QSOs and 
non-BAL compact sources, it is interesting to note in Figure \ref{percentile} that there might be a small deficit 
of steep sources among BAL QSOs with spectral index $\sim -$0.7, compared to the compact sources in 
\cite{Vigotti97}. This effect is more marked in the upper panel, but also insinuated in the lower panel where, 
nevertheless, the number of objects is significantly smaller.

%
%
\subsection{Polarisation}

\begin{table*}
 \begin{minipage}{180mm}
 \centering
  \caption{Degree of polarisation, $m_{\nu}$ (in percentage), at several frequencies, $\nu$ (in GHz), for the RBQ sample. 
  Most of the values are 3-$\sigma$ upper limits and
  those showing a measured value can have a relatively large error but come from an at least 3-sigma detection in Stokes Q
  or Stokes U. The BAL QSOs 1159+01 and 1624+37 presented separately (see Table \protect\ref{Pol1159} in this paper and 
  Table 2 from \protect\citealt{Benn}) are the only two sources with significantly detected polarisation.}\label{pol}
\begin{tabular}{crrrrrrrrrr}
\hline
\multicolumn{1}{c}{ID}&  
\multicolumn{1}{c}{$m_{1.4}^{a}$} &  
\multicolumn{1}{c}{$m_{2.6}$} & 
\multicolumn{1}{c}{$m_{4.8}$} & 
\multicolumn{1}{c}{$m_{8.3}$} & 
\multicolumn{1}{c}{$m_{8.4}$} & 
\multicolumn{1}{c}{$m_{10.5}$}& 
\multicolumn{1}{c}{$m_{15}$}  & 
\multicolumn{1}{c}{$m_{22}$}  &  
\multicolumn{1}{c}{$m_{43}$}\\
\multicolumn{1}{c}{(1)}	&
\multicolumn{1}{c}{(2)}	&
\multicolumn{1}{c}{(3)}	&  
\multicolumn{1}{c}{(4)}	& 
\multicolumn{1}{c}{(5)}	& 
\multicolumn{1}{c}{(6)}	&  
\multicolumn{1}{c}{(7)}	&
\multicolumn{1}{c}{(8)}	& 
\multicolumn{1}{c}{(9)}	& 
\multicolumn{1}{c}{(10)}\\
\hline
0039$-$00& $<$ 1.59 &    $-$   & $<$ 3.8 & $<$ 14.5  & $<$  0.67  & $<$24.8  & $<$10.5	& $<$14.7    & $<$ 8.3      \\
0135$-$02& $<$ 1.26 &    $-$   & $<$ 3.1 & $<$  5.0  & $<$  0.87  & $<$ 8.6  & 3.9$\pm$4.0& $<$ 1.7    & $<$ 7.4    \\
0256$-$01& $<$ 1.41 &    $-$   & $<$ 4.5 & $<$ 26.4  & $<$  3.9	& $<$16.4  & $<$ 1.7	& $<$22.7    & $<$ 7.2      \\
0728+40  & $<$ 1.17 &    $-$   & $<$17.5 & $<$ 30.7  & $<$ 10.1   &  $-$     & $<$10.9	& $<$ 5.1    &   $-$	    \\
0837+36  &    $-$   & $<$ 27.9 & $<$ 2.2 & $<$  6.9  & $<$  0.6   & $<$16.4  & $<$27.8	& $<$15.2    & $<$10.1      \\
0957+23  & $<$ 1.23 &    $-$   & $<$ 0.7 & $<$  1.8  & $<$  0.5   & $<$ 3.4  & $<$ 0.9	& 2.0$\pm$1.4& $<$ 5.6	    \\
1053$-$00& $<$ 1.29 &    $-$   & $<$ 4.2 & $<$ 18.1  & 1.3$\pm$1.0& $<$20.8  & $<$12.8	& $<$ 7.3    & $<$22.1      \\
1213+01  & $<$ 1.29 &    $-$   & $<$ 5.4 & $<$ 13.7  & $<$  0.5   & $<$37.1  & $<$19.0	& $<$ 9.9    & $<$ 7.3      \\
1228$-$01& $<$ 1.35 &    $-$   & $<$ 4.3 & $<$ 10.7  & $<$  0.5   & $<$24.6  & $<$10.9	& $<$ 1.0    & $<$23.7      \\
1312+23  & $<$ 1.17 &    $-$   & $<$ 2.5 & $<$  8.2  & 2.7$\pm$0.7& $<$16.2  & $<$ 1.5	& 3.6$\pm$2.3& $<$ 6.5	    \\
1413+42  & $<$ 1.26 &    $-$   & $<$ 6.7 & $<$  8.9  & $<$  0.5   & $<$15.5  & $<$10.2	& $<$ 1.1    & $<$ 9.5      \\
1603+30  & $<$ 1.32 & $<$ 19.3 & $<$ 2.4 & $<$  5.8  & 1.0$\pm$1.0& $<$14.0  & $<$11.2	& $<$ 0.8    & $<$19.3      \\
1625+48  & $<$ 1.08 & $<$ 24.0 & $<$ 6.4 & $<$  7.4  & $<$  4.9   & $<$16.6  & $<$19.9	& $<$19.8    & 17.6$\pm$14.0\\
\hline
\end{tabular}
\end{minipage}
${}^{a}$ Degree of linear polarisation at 1.4 GHz extracted from the NRAO VLA Sky Survey, NVSS \citep{Condon}
\end{table*}

The RBQ sample was searched in the NVSS database looking for polarised emission and only one BAL QSO, 1159+01, 
shows a high amount of linear polarisation at 1.4 GHz. For the rest of the sources only upper limits could be 
established typically below 1.5 per cent at this frequency.

At higher frequencies and up to 43 GHz we have measured the $S_{Q}$ and $S_{U}$ parameters as explained in 
Section 3, in order to obtain the degree of linear polarisation, $m$, and the polarisation angle $\chi$. 
Table \ref{pol} shows the fractional polarised intensity, mostly upper limits, for the BAL QSOs in the RBQ 
sample. The only sources which are clearly polarised showing significant $m$ at various 
frequencies are 1159+01 and 1624+37. These sources are presented in different tables (Table \ref{Pol1159} in 
this paper and Table 2 from \citealt{Benn}). These two are the only bright sources displaying relatively high 
(i.e., $>$10 percent) polarised intensity at some frequency, having 1159+01 m$_{1.4}$=15 per cent and 1624+37 
with m$_{10.5}$=11 per cent. Apart from these, also the weak source 1625+48 with only 1.7 mJy at 43 GHz shows 
m$_{43} \sim$18 per cent but this measurement is based on a $\gtrsim$ 3-$\sigma$ detection in S$_{U}$ and the 
uncertainty in this measurement is high. From the remaining sources in Table \ref{pol}, 1312+23 shows significant 
polarisation at 8.4 and 22 GHz, and there are 5 more sources with significant detections at only one frequency, 
in all cases below 4 per cent. 

It should be noted from Table \ref{pol} that the most stringent upper limits are given by the VLA observations 
at 8.4 GHz. At this frequency the sources are bright enough to have reliable detections and the relatively low
noise achieved makes possible the determination of reasonable upper limits for the non-detections. At higher 
frequencies this is more difficult because the total power spectra drops down to fainter levels. 

At 8.4 GHz, one third of the sample (i.e, 5 out of 15) show at least a 3-$\sigma$ detection in $S_{Q}$, $S_{U}$ 
or both. From these 5 BAL QSOs only 1624+37 is strongly polarised showing 6.5 per cent of linearly polarised flux
density \citep{Benn}. The fractional polarised intensity in the other four objects (1053$-$00, 1159+01, 1312+23 
and 1603+30) is below 3 per cent. The upper limits for most of the 10 undetected sources are below 1 per cent
indicating strong depolarisation.

For 1159+01 and 1624+37 it is possible the determination of the Rotation Measure, RM, ($\chi=\chi_{0}+RM \cdot 
\lambda^{2}$) fitting a slope to the polarisation angles $\chi_{i}$ as a function of the square of the observed 
wavelenghts. \cite{Benn} already reported for 1624+37 a Rotation Measure of $-$990 $\pm$ 30 rad m$^{-2}$, which 
in the rest-frame translates into the extremely high intrinsic Rotation Measure, RM = -18350 $\pm$ 570 rad m$^{-2}$.
For 1159+01 we have fitted the angles $\chi_{i}$ in Table \ref{Pol1159}. Our observations give a poor upper-limit 
at 2.65 GHz, but an additional measurement at this frequency has been obtained by \cite{Simard}. We get a best fit 
with a Rotation Measure RM=($-$72.1 $\pm$ 1.4) rad m$^{-2}$ and an intrinsic polarisation angle 
$\chi_{0}=-$24 $\pm$ 3 deg. This observed value should be corrected by the RM introduced by our own Galaxy, which 
is quite difficult to estimate. To have a rough estimate we have inspected the all-sky map presented by 
\cite{Wielebinski} where the RM of 976 extragalactic sources are plotted. The neighbouring sources around the
position of 1159+01 have small (i.e., $|$RM$|<$30 rad m$^{-2}$) Rotation Measures, which suggests that the 
Galactic contribution might be smaller than the determined value. Assuming no Galactic correction, the fitted 
Rotation Measure brought to the rest-frame would be higher by a factor (1+z)$^{2}$ becoming 
RM=(644 $\pm$ 12) rad m$^{-2}$.  

\begin{table}
\begin{minipage}{72mm}
 \centering
  \caption{Polarisation properties of BAL QSO 1159+01. The second measurement at 2.65 GHz 
  comes from \protect\cite{Simard}}\label{Pol1159}
  \begin{tabular}{@{}cccc@{}}
\hline \hline
     Frequency     & Telescope &    $m$    &  $\chi$ \\
    (GHz)          &	        &(per cent) &   (deg) \\
\hline
 1.40 & NVSS	    & 15.03 $\pm$ 0.44 & $-$31.1 $\pm$ 0.6 \\
 2.65 & Effelsberg  & $<$16	       &     $-$  	 \\
 2.65 & $-$         &  3.1  $\pm$ 1.9  &    111 $\pm$ 17 \\
 4.85 & Effelsberg  &  2.1  $\pm$ 0.2  &  $-$42 $\pm$ 6	 \\
 8.35 & Effelsberg  &  0.95 $\pm$ 0.05 &  $-$29 $\pm$ 5	 \\
 8.45 & VLA	    &  0.9  $\pm$ 0.4  &  $-$29 $\pm$ 7	 \\
10.5  & Effelsberg  &  $<$1.2	       &   $-$  	 \\
15.0  & VLA	    &  $<$1.2          &   $-$           \\
22.5  & VLA	    &  $<$0.4	       &   $-$  	 \\
43.5  & VLA	    &  $<$1.2	       &   $-$  	 \\
\hline 
\end{tabular}
\end{minipage}
\end{table}

\begin{figure}
 \begin{minipage}{72mm}
  \includegraphics[width=84mm]{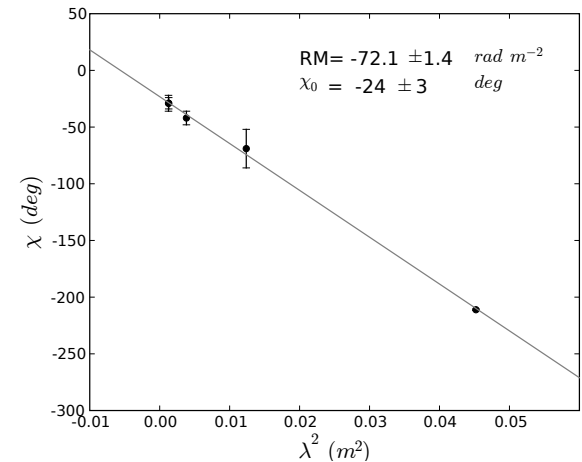}
  \caption{Determination of the RM for 1159+01.}\label{sizes}
\end{minipage}
\end{figure}

\section{Discussion}\label{sec6}

The morphologies and dimensions of the radio sources in BAL QSOs seem to be the keys to understand the 
orientation and evolutionary status of these sources. There is evidence that BAL QSOs are associated with 
compact radio sources, which are supposed to constitute a high fraction of the young population of radio 
sources, like e.g., CSS or GPS sources. \cite{Becker1} noted that 90 per cent of their sample of radio-loud 
BAL QSOs (extracted from the FBQS survey) present point-like structure at the resolution of 5$\arcsec$ in 
contrast to the diversity of both point-like and extended sources they found within the whole population 
of FBQS quasars. In fact, only a few BAL QSOs are known to have an extended FR II radio structure 
\citep{Gregg2}. Our 22-GHz observations of the RBQ sample confirm this tendency, because most sources 
appear unresolved at all frequencies constraining their apparent dimensions up to $\lesssim$0.1 arcsec. 
This translates into projected linear sizes (LS) $\lesssim$ 1 kpc which are typical of GPS/CSS sources. 
It could be argued that an extended component with a steep spectrum might be present, observable only 
at lower frequencies, and thus being the source sizes larger than those found at high frequencies. 
This hypothesis is \emph{a priori} no supported by the fact that most sources are still point-like at 
1.4 GHz as can be seen in the FIRST maps, with the exception of 1053$-$00. 

The radio spectra in Figure \ref{synage_fits} also display the typical convex shape of GPS-like sources, in most 
cases with a peak at $\sim$ 0.5$-$10 GHz suggested by at least one or two points, or by upper limits at MHz 
frequencies. One concern about this result is that FIRST or WENSS flux densities were observed at different 
epochs and variability might be an issue, although noticeable flux density variations are likely to be stronger
in the optically thin region of the spectrum. However, relatively deep integrations at MHz frequencies would be 
desirable to better constrain the shape of the spectra at low frequency. In addition, a project with VLBI 
multi-frequency observations and the analysis on the pc-scale properties of several BAL QSOs of the RBQ 
sample has been started by us, and results will be presented in a subsequent paper.

\begin{figure*}
 \begin{minipage}{174mm}
\centering
  \includegraphics[width=152mm]{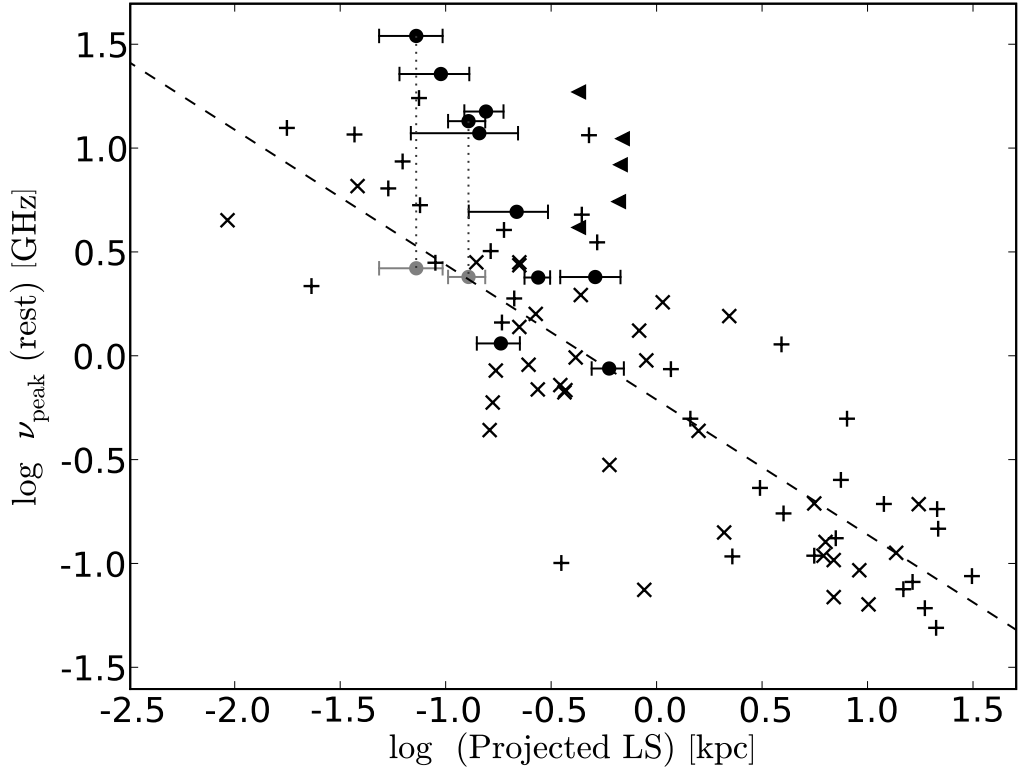}
  \caption{Rest-frame turnover frequency versus projected linear size relation for samples of CSS and GPS 
  sources (adapted from \protect\citealt{ODea2}). Galaxies are represented by the symbol '$\times$', quasars 
  by the simbol '+' and the RBQ sample is plotted as solid circles. Upper limits for those BAL QSOs unresolved 
  at 22 or 43 GHz are marked by triangles. The dashed line represent the anticorrelation found by 
  \protect\cite{ODea2}: log $\nu_{peak}$ = $-$0.21$-$0.65 log LS. In those two resolved sources with two peaks 
  in the spectrum (1413+43 and 1603+30) the grey points indicate the new location in the plot when the lower 
  frequency peak is considered.
  }\label{nu_vs_ls_relation}
\end{minipage}
\end{figure*}

If radio sources associated to BAL QSOs are actually CSS/GPS sources, they should follow the anticorrelation
between intrinsic turnover frequency and projected linear size found by \cite{Fanti90} for CSS sources. 
This was, for instance, quantified by \cite{ODea2} in a combined sample of CSS and GPS sources compiled 
by \cite{Fanti90} and \cite{Stanghellini}, respectively. This relationship was found to be valid for both galaxies 
and quasars. The anticorrelation suggests that the mechanism producing the turnover simply depends on the source 
size. Figure \ref{nu_vs_ls_relation} shows the relationship between intrinsic $\nu_{peak}$ and projected LS for 
the samples of CSS and CPS sources mentioned before, and we have also plotted for comparison our sample of 
radio-loud BAL QSOs. The turnover frequencies for the BAL QSOs appear in Table \ref{listalpha} while projected 
linear sizes have been extracted from the deconvolved sizes in Table \ref{dec_sizes}.

A group of 5 BAL QSOs match quite well the distribution of CSS/GPS sources, and they all show simple convex 
spectra, i.e., 0728+40, 1213+01, 1312+23, 1624+37 and 1625+48. The three additional BAL QSOs displaying simple 
convex spectrum, 0039$-$00, 0256$-$01 and 0957+23 are unresolved in our observations and their position can only
be constrained by upper limits. Nevertheless these 8 sources with simple complex spectrum are consistent with
the correlation. 

From the remaining 7 objects, 5 are resolved in our observations (0135$-$02, 0837+36, 1228$-$01, 1413+42 and 
1603+30) and they are located above the cloud of CSS/GPS sources. All of them have in common a complex spectrum 
with indications of a double component. In these sources the turnover has been determined using the high frequency
component but it is possible in the spectra of 1413+42 and 1603+30 to fit a second peak at lower frequencies 
(at 0.7 and 0.8 GHz respectively). A second peak at low frequency could also be fitted to 1159+01 but we will 
not do that because being this source unresolved we cannot give the exact location in the diagram. When considering
the low frequency peaks of 1413+42 and 1603+30 in the diagram, these two sources also agree very well with the 
anticorrelation as shown in Figure \ref{nu_vs_ls_relation}.  This better agreement 
could mean that the high-frequency part of these spectra could be dominated by the emission coming from a compact 
and active (possibly beamed) region smaller than the entire source, like a hot-spot. Unfortunately we do not have 
enough data to check whether this could be the case also for 0135-02, 0837+36 and 1228-01. Low-frequency data will 
be important to sample this part of the spectra to search for a possible second peak. Again, VLBI observations are 
crucial to further investigate this interpretation.

In fact, it has been shown that some GPS sources associated to optical QSOs can be just normal flat-spectrum 
quasars for which a jet/knot/hot-spot dominates the spectrum, which then adopts the characteristic convex shape 
of a GPS source, as discussed e.g. by \cite{Snellen}. Recently \cite{Stanghellini05} and \cite{Orienti} have 
studied VLBI samples of GPS sources and High Frequency Peakers (HFP, 5 GHz $< \nu_{\rm{pk}} <$ 10 GHz) confirming 
this scenario. In fact, these ``contaminant'' quasars seem to be present, even in high percentages, in many 
well-defined samples of GPS sources \citep{Torniainen05}. Two characteristics of ``genuine'' young GPS radio 
sources are low polarisation and low variability \citep{ODea}, while flat-spectrum quasars show higher 
polarisation and variability. 

\cite{Torniainen05}, who have used an extensive database of sources with multifrequency data from several 
long-term variability programs, define as ``bona fide'' GPS sources those with a maximum variability 
$Var_{\Delta S}<$3. From our comparison in Section \ref{sec4} which is, of course, based on only two 
epochs we see that many sources do not show significant variations at 1.4 GHz, and 3 out of 5 inspected at
8.4 GHz show significant but not very strong variations. We cannot thus exclude these sources 
from the candidates to young sources. As also shown in \cite{Torniainen05} the maximum variations 
are found at high frequencies, so multi-frequency multi-epoch observations of the RBQ sample will be very 
interesting for a more precise variability study. 

Compact GPS sources are supposed to be completely embedded in their narrow line region. On this basis, 
\citep{Cotton} proposed the existence of a typical frequency-dependent scale, below which the GPS source 
becomes completely depolarised, the so-called ``Cotton effect''. \cite{Fanti04} and later 
\cite{Rossetti} have confirmed this behaviour analysing the polarisation properties of the B3-VLA CSS 
sample \citep{Fanti01}, and they propose models to describe the depolarising screen. The BAL QSOs in the 
RBQ sample have probably physical sizes of $\lesssim$ 1 kpc and if Faraday effects are important they should be 
completely depolarised even up to relatively high frequencies. This is the case for all sources except 1159+01
at 1.4 GHz, and also for about 70 per cent of the sample at 8.4 GHz. At these two frequencies 
the upper limits obtained for the degree of linear polarisation are reasonably small. From Table \ref{pol} 
only 1312+23 seems to be weakly polarised at 8.4 and 22 GHz, but no polarisation was found at 15 GHz.

The two exceptions are 1159+01 and 1624+37. The first source is weakly polarised at 8.4 GHz but it is 
worth to note that the degree of linear polarisation increases as frequency decreases, 
becoming strongly polarised at 1.4 GHz. The opposite is expected when a depolarising plasma curtain is 
present. The reason for this anomalous behaviour might be that the emission at low frequencies arises from a 
different component, as suggested by the complex radio spectrum (see Figure \ref{synage_fits}). An 
hypothetical extended structure with a relatively ordered magnetic field would contribute with substantial 
linear polarisation, and this may only be detected at low frequencies if a steep spectral index is assumed 
for the emission of this extended component. 

The moderate observed Rotation Measure of ($-$72.1 $\pm$ 1.4) rad m$^{-2}$ in 1159+01 is consistent 
with the fact that it is strongly polarised in the NVSS. Should the Rotation Measure be much higher, the 
large bandwidth used for the NVSS observations would produce a strong bandwidth depolarisation effect, 
probably hiding any measurable polarised intensity. The extreme Rotation Measure of J1624+37, RM=$-$18350 
rad m$^{-2}$ \citep{Benn} which is the second highest value known among QSOs and the highest among BAL QSOs, 
seems to be an exception and not representative of the BAL QSOs family. 

The median $m_{8.4}$ of the 5 polarised BAL QSOs in the RBQ sample is $\sim$1.3 per cent, but the remaining
10 BAL QSOs show $m_{8.4} \lesssim$1. As a comparison \cite{Saikia} found 
a median polarisation degree at 6 GHz of about 2 per cent in a subsample of quasars extracted from the sample 
of $\sim$400 compact sources of \cite{Perley}. \cite{Saikia} found no significant differences between the median 
value of flat-spectrum cores and CSS quasars whereas that radio sources associated to galaxies or to empty fields 
(no optical identifications) were found to be less polarised at 6 GHz, with a median value of $\sim$0.5 per cent.
More recently, \cite{Stanghellini03} found a mean fractional polarisation of 1.2 and 1.8 per cent for GPS and 
flat-spectrum quasars, respectively, and $m<$ 0.3 per cent for galaxies. 

The sources showing some polarisation in the RBQ sample have probably polarised intensities consistent with 
flat-spectrum quasars. It is also likely that some of them (e.g., 1159+01 or 1603+30) can show polarised emission 
due to Doppler boosted knots in jets, which is again compatible with their multi-component radio spectra peaking 
at high frequencies. However, our whole sample seem to be in better agreement with the GPS class than with the 
class of flat-spectrum quasars, because of the high percentage of unpolarised or weakly polarised sources. 
Of course, to firmly test this hypothesis it will be particularly interesting to do polarimetric multi-frequency 
observations of a radio brighter sample, which is now available after the recent releases of SDSS.

As far as the spectral index distribution is concerned, the different statistical tests suggest that the 
spectral index distributions of BAL and non-BAL QSOs are different. This difference is however due to the fact 
that all BAL QSOs in the sample have compact structures, while non-BAL QSOs have both extended and compact 
morphologies. When comparing the samples of BAL QSOs and compact non-BAL QSOs, no significant differences 
are found in the spectral index distributions. 
Thus, from our comparison no preferred orientation can be attributed to BAL QSOs. 
When larger samples become available, the comparison of spectral indices of BAL and non-BAL QSOs as a function 
of radio power and UV luminosity will be possible, allowing us to look for particular orientations in different 
bins of luminosity, as has been suggested by \cite{Elvis}.

\section{Conclusions}\label{sec7}
\begin{itemize}
\item A sample of 15 BAL QSOs associated with FIRST sources brighter than 15 mJy has been built. Radio continuum 
observations for these sources have been collected from 2.6 up to 43 GHz in full polarisation. Flux densities 
covering this radio range have been presented complemented by archive data at lower frequencies from 74 MHz up
to 1.4 GHz.

\item VLA maps in the most extended configuration show very compact morphologies for most sources at all frequencies 
being unresolved or slightly resolved at 22 GHz with a resolution of 80 mas. These translate into projected linear 
sizes $\lesssim$ 1 kpc, which are the typical sizes of CSS/GPS sources. 

\item The spectra of these sources are typically convex, i.e., they seem to become flat or inverted at MHz frequencies 
probably due to synchrotron self-absorption, while at frequencies higher than 15 GHz they are definitely steeper. 
The spectra typically peak between 1 and 5 GHz in the observer's frame and about 1/3 of the sample shows complex 
spectra suggesting different components.

\item All BAL QSOs in our sample presenting simple convex spectra follow the anticorrelation between projected 
linear size and turnover frequency found for CSS and GPS sources. Two BAL QSOs with complex radio spectra 
are also in agreement with this relationship if we consider the lower frequency peak of their spectra. The higher 
frequency peak might be in this two sources due to emission from a hot-spot or a knot in a jet.

\item A high percentage of the sample does not show significant variability when comparing the flux densities in
2 different epochs. Some BAL QSOs show significant variability at 8.4 GHz but not strong enough to exclude them 
from the candidates to young radio sources. 

\item Most BAL QSOs in our list are not strongly polarized either at 1.4 GHz or at 8.4 GHz, and sensible upper limits
are given at these frequencies. Only two sources show significant polarisation at several frequencies. From these
two, only the unusual BAL 1624+37 has a high Rotation Measures suggesting strong depolarization. The median fractional 
polarisation of the sample is in better agreement to mean values found in CSS/GPS sources, than those found for 
flat-spectrum quasars. 

\item A series of statistical tests have been done to compare the spectral index distribution of BAL and 
non-BAL QSOs finding that these distributions are different. However, no significant differences were found when 
comparing to compact non-BAL QSOs only. This is consistent with BAL QSOs spanning the same range of
orientations as normal quasars with respect to our line of sight to them.

\end{itemize}
\section*{Acknowledgements}

We are grateful to F. Mantovani and M. Orienti for helping us during the observations at the 100-m Effelsberg telescope, and to
A. Kraus for support and help with the calibration of the Effelsberg data. 
The authors acknowledge financial support from the Spanish Ministerio 
de Educaci\'on y Ciencia under project PNAYA2005-00055. This work has benefited from research funding from the European Community's 
sixth Framework Programme under RadioNet R113CT 2003 5058187. This work has been partially based on observations with the 100-m 
telescope of the MPIfR (Max-Planck-Institut f\"ur Radioastronomie) at Effelsberg. The National Radio Astronomy Observatory is a 
facility of the National Science Foundation operated under cooperative agreement by Associated Universities, Inc. This research 
has made use of the NASA/IPAC 
Infrared Science Archive and NASA/IPAC Extragalactic Database (NED) which are both operated by the Jet Propulsion Laboratory, 
California Institute of Technology, under contract with the National Aeronautics and Space Administration. Use has been made
of the Sloan Digital Sky Survey (SDSS) Archive. The SDSS is managed by the Astrophysical Research Consortium (ARC) for the 
participating institutions: The University of Chicago, Fermilab, the Institute for Advanced Study, the Japan Participation
Group, The John Hopkins University, Los Alamos National Laboratory, the Max-Planck-Institute for Astronomy (MPIA), the 
Max-Planck-Institute for Astrophysics (MPA), New Mexico State University, University of Pittsburgh, Princeton University,
the United States Naval Observatory, and the University of Washington.

\label{lastpage}

\begin{thebibliography}{99}

\bibitem[\protect\citeauthoryear{Aller, Aller \& Hughes}{Aller et al.}{2003}]{Aller} Aller M.F., Aller H.D., 
Hughes P.A., 2003, ApJ, 586, 33
\bibitem[\protect\citeauthoryear{Baars et al.}{1977}]{Baars} Baars J.W.M., Genzel
R., Pauliny-Toth I.I.K., Witzel A., 1977, A\&A, 61, 99
\bibitem[\protect\citeauthoryear{Becker, White \& Helfand}{Becker et al.}{1995}]{Becker95} Becker R.H.,
White R.L., Helfand D.J., 1995, ApJ, 450, 559
\bibitem[\protect\citeauthoryear{Becker et al.}{2000}]{Becker1} Becker R.H., White R.L., Gregg M.D., 
Brotherton M.S., Laurent-Muehleisen S.A., Arav N., 2000, ApJ, 538, 72
\bibitem[\protect\citeauthoryear{Becker et al.}{2001}]{Becker2} Becker R.H. et al.,
2001, ApJS, 135, 227
\bibitem[\protect\citeauthoryear{Benn et al.}{2005}]{Benn} Benn C.R., Carballo R., Holt J., Vigotti M., 
Gonz\'alez-Serrano J.I., Mack K.-H., Perley R.A., 2005, MNRAS, 360, 1455
\bibitem[\protect\citeauthoryear{Cohen et al.}{2007}]{Cohen} Cohen A.S., Lane W.M., Cotton W.D., Kassim N.E., 
Lazio T.J.W., Perley R.A., Condon J.J., Erickson W.C., 2007, AJ, 134, 1245
\bibitem[\protect\citeauthoryear{Condon et al.}{1998}]{Condon} Condon J.J., Cotton W.D., Greisen E.W., 
Yin Q.F., Perley R.A., Taylor G.B., Broderick J.J., 1998, AJ, 115, 1693
\bibitem[\protect\citeauthoryear{Cotton et al.}{2003}]{Cotton} Cotton W.D. et al.,  
2003, PASA, 20, 12
\bibitem[\protect\citeauthoryear{Dai, Shankar \& Sivakoff}{Dai et al.}{2008}]{Dai} Dai X., Shankar F., Sivakoff G.R.,
2008, ApJ, 672, 108
\bibitem[\protect\citeauthoryear{Dallacasa et al.}{2000}]{Dallacasa} Dallacasa D., Stanghellini C., Centoza M.,
Fanti R., 2000, A\&A, 363, 887
\bibitem[\protect\citeauthoryear{de Vries, Becker \& White}{de Vries et al.}{2006}]{DeVries} De Vries W.H., 
Becker R.H., White R.L., 2006, AJ, 131, 666
\bibitem[\protect\citeauthoryear{Douglas et al.}{1996}]{Douglas} Douglas J.N., Bash F.N., Bozyan F.A.,
Wolfe C., 1996, AJ, 111, 1945
\bibitem[\protect\citeauthoryear{Djorgovski et al.}{1990}]{Djorgovski} Djorgovski S., Thompson D.J., 
Vigotti M., Grueff G., 1990, PASP 102, 113
\bibitem[\protect\citeauthoryear{Elvis}{2000}]{Elvis} Elvis M., 2000, ApJ, 545, 63
\bibitem[\protect\citeauthoryear{Fanti et al.}{1990}]{Fanti90} Fanti R., Fanti C., Schilizzi R.T., Spencer R.E., 
Rendong N., Parma P., van Breugel W.J.M., Venturi T., 1990, A\&A, 231, 333
\bibitem[\protect\citeauthoryear{Fanti et al.}{1995}]{Fanti} Fanti C., Fanti R., Dallacasa D.,
Schilizzi R.T., Spencer R.E., Stanghellini C., 1995, A\&A, 302, 317
\bibitem[\protect\citeauthoryear{Fanti et al.}{2001}]{Fanti01} Fanti C., Pozzi F., Dallacasa D., Fanti R., 
Gregorini L., Stanghellini C., Vigotti M., 2001, A\&A, 369, 380
\bibitem[\protect\citeauthoryear{Fanti et al.}{2004}]{Fanti04} Fanti C. et al., 
2004, A\&A, 427, 465
\bibitem[\protect\citeauthoryear{Ficarra, Grueff \& Tomassetti et al.}{Ficarra et al.}{1985}]{Ficarra} Ficarra A., 
Grueff G., Tomassetti G., 1985, A\&AS, 59, 255
\bibitem[\protect\citeauthoryear{Ganguly et al.}{2007}]{Ganguly} Ganguly R., Brotherton M.S., 
Cales S., Scoggins B., Shang Z., Vestergaard M., 2007, ApJ, 665, 990
\bibitem[\protect\citeauthoryear{Ganguly \& Brotherton}{2008}]{Ganguly2} Ganguly R., Brotherton M.S., 
2008, ApJ, 672, 102
\bibitem[\protect\citeauthoryear{Gregorini et al.}{1998}]{Gregorini} Gregorini L., Vigotti M., 
Mack K.-H., Z\"onnchen J., Klein U., 1998, A\&AS, 133, 129
\bibitem[\protect\citeauthoryear{Gregg et al.}{1996}]{Gregg} Gregg M.D., Becker R.H., 
White R.L., Helfand D.J., McMahon R.G., Hook I.M., 1996, AJ, 112, 407
\bibitem[\protect\citeauthoryear{Gregg, Becker \& de Vries}{Gregg et al.}{2006}]{Gregg2} Gregg M.D., Becker R.H., 
de Vries W., 2006, ApJ, 641, 210
\bibitem[\protect\citeauthoryear{Ghosh \& Punsly}{2007}]{Ghosh} Ghosh K.K., Punsly B., 2007, 
ApJ, 661L, 139
\bibitem[\protect\citeauthoryear{Hall et al.}{2002}]{Hall} Hall P., et al., 2002, ApJS, 141, 267
\bibitem[\protect\citeauthoryear{Hamann \& Sabra}{2004}]{Hamann} Hamann F., Sabra B., 2003, in
Richards G.T., Hall P.B., eds, ASP Conf. Ser. Vol. 311, AGN Physics with the Sloan Digital Sky Survey.
Astron. Soc. Pac., San Francisco, p. 203
\bibitem[\protect\citeauthoryear{Henstock et al.}{1997}]{Henstock} Henstock D. R., Browne I.W.A., 
Wilkinson P.N., McMahon R.G., 1997, MNRAS 290, 380
\bibitem[\protect\citeauthoryear{Hewett \& Foltz}{2003}]{Hewett} Hewett P.C., Foltz C.B., 2003, 
AJ, 125, 1784
\bibitem[\protect\citeauthoryear{Holt et al.}{2004}]{Holt} Holt J., Benn C.R., Vigotti M., Pedani M., Carballo R., 
Gonz\'{a}lez-Serrano J.I., Mack K.-H., Garc\'{i}a B., 2004, MNRAS, 348, 857
\bibitem[\protect\citeauthoryear{Holt et al.}{2008}]{Holt2} Holt J., Tadhunter C.N., Morganti R., 
2008, preprint (arXiv:0802.1444) Accepted for publication in MNRAS
\bibitem[\protect\citeauthoryear{Hopkins et al.}{2005}]{Hopkins} Hopkins P.F., Hernquist L., Cox T.J.,
Di Matteo T., Martini P., Robertson B., Springer V., 2005, ApJ, 630, 705
\bibitem[\protect\citeauthoryear{Jiang \& Wang}{2003}]{Jiang} Jiang D.R., Wang T.G., 2003, A\&A, 397, L13
\bibitem[\protect\citeauthoryear{Klein et al.}{2003}]{Klein} Klein U., Mack K.-H., Gregorini L., Vigotti M., 
2003, A\&A, 406, 579
\bibitem[\protect\citeauthoryear{Kunert-Bajraszewska \& Marecki}{2007}]{Magda} Kunert-Bajraszewska M., 
Marecki A., 2007, A\&A, 469, 437
\bibitem[\protect\citeauthoryear{Lahulla et al.}{1991}]{Lahulla} Lahulla J.F., Merighi R., Vettolani G., 
Vigotti M., 1991, A\&AS, 88, 527
\bibitem[\protect\citeauthoryear{Mack et al.}{2005}]{Mack} Mack K.-H., Vigotti, M., Gregorini L., Klein U.,
Tschager W., Schilizzi R.T., Snellen I.A.G., 2005, A\&A, 435, 863
\bibitem[\protect\citeauthoryear{Menou et al.}{2001}]{Menou} Menou K. et al., 2001, ApJ, 561, 645
\bibitem[\protect\citeauthoryear{Miley}{1980}]{Miley} Miley G., 1980, ARA\&A, 18, 165
\bibitem[\protect\citeauthoryear{O'Dea}{1998}]{ODea} O'Dea C.P., 1998, PASP, 110, 493
\bibitem[\protect\citeauthoryear{O'Dea \& Baum}{1997}]{ODea2} O'Dea C.P, Baum S.A., 1997, AJ, 113, 148.
\bibitem[\protect\citeauthoryear{Orienti et al.}{2006}]{Orienti} Orienti M., Dallacasa D., 
Tinti S., Stanghellini C., 2006, A\&A, 450, 959
\bibitem[\protect\citeauthoryear{Orr \& Browne}{1982}]{Orr} Orr M.J.L., Browne I.W.A., 1982, MNRAS, 200, 1067
\bibitem[\protect\citeauthoryear{Osmer, Porter \& Green}{Osmer et al.}{1994}]{Osmer} Osmer P.S., Porter A.C., 
Green R.F., 1994, ApJ, 436, 678
\bibitem[\protect\citeauthoryear{Perley}{1982}]{Perley} Perley R.A., 1982, AJ, 87, 859
\bibitem[\protect\citeauthoryear{Rengelink et al.}{1997}]{Rengelink} Rengelink R.B., Tang Y., de Bruyn A.G., 
Miley G.K., Bremer  M.N., R\"ottgering  H.J.A., Bremer M.A.R., 1997, A\&AS, 124, 259
\bibitem[\protect\citeauthoryear{Rei\-chard et al.}{2003a}]{Reichard} Reichard T.A. et al.,
2003a, AJ, 125, 1171
\bibitem[\protect\citeauthoryear{Rei\-chard et al.}{2003b}]{Reichard2} Reichard T.A. et al.,
2003b, AJ, 126, 2594
\bibitem[\protect\citeauthoryear{Rossetti et al.}{2008}]{Rossetti} Rossetti A., Dallacasa D., Fanti C., Fanti R., 
Mack K.-H., 2008, A\&A, in press
\bibitem[\protect\citeauthoryear{Saikia, Singal \& Cornwell}{Saikia et al.}{1987}]{Saikia} Saikia D.J., 
Singal A.K., Cornwell T.J., 1987, MNRAS 234, 379
\bibitem[\protect\citeauthoryear{Schlegel, Finkbeiner \& Davis}{Schlegel et al.}{1998}]{Schlegel} Schlegel D.J., 
Finkbeiner D.P., Davis M., 1998, ApJ, 500, 525
\bibitem[\protect\citeauthoryear{Schneider et al.}{2005}]{Schneider05} Schneider D.P. et al, 
2005, AJ, 130, 367
\bibitem[\protect\citeauthoryear{Schneider et al.}{2007}]{Schneider07} Schneider D.P. et al, 
2007, AJ, 134, 102
\bibitem[\protect\citeauthoryear{Simard-Normandin, Kronberg \& Button}{Simard-Normandin et al.}{1981}]{Simard} 
Simard-Normandin M., Kronberg P.P., Button S., 1981, ApJS, 46, 239
\bibitem[\protect\citeauthoryear{Snellen et al.}{1999}]{Snellen} Snellen I.A.G., Schilizzi R.T.,
Bremer M.N., Miley G.K., de Bruyn A.G., R\"ottgering H.J.A., 1999, MNRAS, 307, 149
\bibitem[\protect\citeauthoryear{Stanghellini}{1992}]{Stanghellini} Stanghellini C., 1992, Ph.D. thesis, 
University of Bologna
\bibitem[\protect\citeauthoryear{Stanghellini}{2003}]{Stanghellini03} Stanghellini C., 2003, PASA, 20, 118
\bibitem[\protect\citeauthoryear{Stanghellini et al.}{2005}]{Stanghellini05} Stanghellini C., O'Dea C.P., 
Dallacasa D., Cassaro P., Baum S.A., Fanti R., Fanti C., 2005, A\&A, 443, 891
\bibitem[\protect\citeauthoryear{Stickel \& K\"uhr}{1993}]{Stickel} Stickel M., K\"uhr H., 1993, A\&AS, 101, 521
\bibitem[\protect\citeauthoryear{Stocke et al.}{1992}]{Stocke} Stocke J.T., Morris S.L., Weymann R.J., Foltz C.B., 
1992, ApJ, 396, 487
\bibitem[\protect\citeauthoryear{Tabara \& Inoue}{1980}]{Tabara} Tabara H., Inoue M., 1980, A\&AS, 39, 379
\bibitem[\protect\citeauthoryear{Torniainen et al.}{2005}]{Torniainen05} Torniainen H., Tornikoski M., 
Ter\"asranta H., Aller M.F., Aller H.D., 2005, A\&A, 435, 839
\bibitem[\protect\citeauthoryear{Torniainen et al.}{2007}]{Torniainen07} Torniainen H., Tornikoski M., 
L\"ahteenm\"aki A., Aller M.F., Aller H.D., Mingaliev M.G., 2007, A\&A, 469, 451
\bibitem[\protect\citeauthoryear{Trump et al.}{2006}]{Trump} Trump J.R. et al., 2006, ApJS, 165, 1
\bibitem[\protect\citeauthoryear{Vigotti et al.}{1989}]{Vigotti89} Vigotti M., Grueff G.,
Perley R., Clark B.G., Bridle A.H., 1989, AJ, 98, 419
\bibitem[\protect\citeauthoryear{Vi\-go\-tti et al.}{1997}]{Vigotti97} Vigotti M., Vettolani G.,
Merighi R., Lahulla J.F., Pedani M., 1997, A\&A, 123, 219
\bibitem[\protect\citeauthoryear{Vi\-go\-tti et al.}{1999}]{Vigotti99} Vigotti M., Gregorini L.,
Klein U., Mack K.-H., 1999, A\&AS, 139, 359
\bibitem[\protect\citeauthoryear{Weymann et al.}{1991}]{Weymann} Weymann R.J., Morris S.L., Foltz C.B., 
Hewett P.C., 1991, ApJ, 373, 23
\bibitem[\protect\citeauthoryear{White et al.}{1997}]{White} White R.L., Becker R.H., Helfand D.J., 
Gregg M.D., 1997, ApJ, 475, 479
\bibitem[\protect\citeauthoryear{Wielebinski \& Krause}{1993}]{Wielebinski} Wielebinski R., Krause F., 
1993, A\&AR, 4, 4
\bibitem[\protect\citeauthoryear{Zhou et al.}{2006}]{Zhou} Zhou H., Wang T., Wang H., Wang J., Yuan W., 
Lu Y., 2006, ApJ, 639, 716
\end{thebibliography}
\end{document}